\documentclass[aps,prc,twocolumn,floatfix,showpacs,preprintnumbers,amsmath,amssymb,superscriptaddress]{revtex4}
\usepackage{hyperref}
\usepackage{latexsym}
\usepackage{verbatim}
\usepackage{graphics}
\usepackage{caption}
\usepackage{subfigure}
\usepackage{setspace}
\usepackage{graphicx,amsmath,color}

\definecolor{lightred}{rgb}{1,0.5,0.5}
\definecolor{lightgreen}{rgb}{0.5,1,0.5}
\definecolor{darkgreen}{rgb}{0.5,0.9,0.5}

\begin{document}
\preprint{}

\title{$\phi$-meson photoproduction on Hydrogen in the neutral decay mode}

\newcommand*{\ODU}{Old Dominion University, Norfolk, Virginia 23529}
\affiliation{\ODU}
\newcommand*{\BOCHUM}{Institut f\"ur Theoretische Physik II,
Ruhr--Universit\"at Bochum, D--44780 Bochum, Germany}
\newcommand*{\PNPI}{Petersburg Nuclear Physics
Institute, Gatchina, St.\ Petersburg 188300, Russia}
\newcommand*{\KYUNGPOOK} {Kyungpook National University, 702-701, Daegu, Republic of Korea}
\newcommand*{\INR}{Institute for Nuclear Research, 117312, Moscow, Russia}
\newcommand*{\GWU}{The George Washington University, Washington, DC 20052}
\newcommand*{\CUA}{Catholic University of America, Washington, DC 20064} 
\newcommand*{\JLAB}{Thomas Jefferson National Accelerator Facility, Newport News, Virginia 23606}
\newcommand*{\UVA}{University of Virginia, Charlottesville, Virginia 22904}
\newcommand*{\CMU}{Carnegie Mellon University, Pittsburg, PA 15213}

\author {H.~Seraydaryan}
\affiliation{\ODU}
\author {M.J.~Amaryan}
\email{mamaryan@odu.edu}
\thanks{Corresponding author.}
\affiliation{\ODU}
\author {G.~Gavalian}
\affiliation{\ODU}
\author {H.~Baghdasaryan}
\affiliation{\UVA}
\author {L.~Weinstein}
\affiliation{\ODU}


\newcommand*{\ANL}{Argonne National Laboratory, Argonne, Illinois 60439}
\newcommand*{\ASU}{Arizona State University, Tempe, Arizona 85287-1504}
\newcommand*{\CSUDH}{California State University, Dominguez Hills, Carson, CA 90747}
\newcommand*{\CANISIUS}{Canisius College, Buffalo, NY}
\newcommand*{\SACLAY}{CEA, Centre de Saclay, Irfu/Service de Physique Nucl\'eaire, 91191 Gif-sur-Yvette, France}
\newcommand*{\CNU}{Christopher Newport University, Newport News, Virginia 23606}
\newcommand*{\UCONN}{University of Connecticut, Storrs, Connecticut 06269}
\newcommand*{\EDINBURGH}{University of Edinburgh, Edinburgh EH9 3JZ, United Kingdom}
\newcommand*{\FU}{Fairfield University, Fairfield CT 06824}
\newcommand*{\FIU}{Florida International University, Miami, Florida 33199}
\newcommand*{\FSU}{Florida State University, Tallahassee, Florida 32306}
\newcommand*{\GWUI}{The George Washington University, Washington, DC 20052}
\newcommand*{\ISU}{Idaho State University, Pocatello, Idaho 83209}
\newcommand*{\INFNFE}{INFN, Sezione di Ferrara, 44100 Ferrara, Italy}
\newcommand*{\INFNFR}{INFN, Laboratori Nazionali di Frascati, 00044 Frascati, Italy}
\newcommand*{\INFNGE}{INFN, Sezione di Genova, 16146 Genova, Italy}
\newcommand*{\INFNRO}{INFN, Sezione di Roma Tor Vergata, 00133 Rome, Italy}
\newcommand*{\ORSAY}{Institut de Physique Nucl\'eaire ORSAY, Orsay, France}
\newcommand*{\ITEP}{Institute of Theoretical and Experimental Physics, Moscow, 117259, Russia}
\newcommand*{\JMU}{James Madison University, Harrisonburg, Virginia 22807}
\newcommand*{\KNU}{Kyungpook National University, Daegu 702-701, Republic of Korea}
\newcommand*{\LPSC}{LPSC, Universite Joseph Fourier, CNRS/IN2P3, INPG, Grenoble, France
}
\newcommand*{\UNH}{University of New Hampshire, Durham, New Hampshire 03824-3568}
\newcommand*{\NSU}{Norfolk State University, Norfolk, Virginia 23504}
\newcommand*{\OHIOU}{Ohio University, Athens, Ohio  45701}
\newcommand*{\RPI}{Rensselaer Polytechnic Institute, Troy, New York 12180-3590}
\newcommand*{\ROMAII}{Universita' di Roma Tor Vergata, 00133 Rome Italy}
\newcommand*{\MSU}{Skobeltsyn Nuclear Physics Institute, 119899 Moscow, Russia}
\newcommand*{\SCAROLINA}{University of South Carolina, Columbia, South Carolina 29208}
\newcommand*{\UNIONC}{Union College, Schenectady, NY 12308}
\newcommand*{\UTFSM}{Universidad T\'{e}cnica Federico Santa Mar\'{i}a, Casilla 110-V Valpara\'{i}so, Chile}
\newcommand*{\GLASGOW}{University of Glasgow, Glasgow G12 8QQ, United Kingdom}
\newcommand*{\VIRGINIA}{University of Virginia, Charlottesville, Virginia 22901}
\newcommand*{\WM}{College of William and Mary, Williamsburg, Virginia 23187-8795}
\newcommand*{\YEREVAN}{Yerevan Physics Institute, 375036 Yerevan, Armenia}
\newcommand*{\NOWCNU}{Christopher Newport University, Newport News, Virginia 23606}
\newcommand*{\NOWLANL}{Los Alamos National Laboratory, Los Alamos, NM 87544 USA}
\newcommand*{\NOWMSU}{Skobeltsyn Nuclear Physics Institute, 119899 Moscow, Russia}
\newcommand*{\NOWORSAY}{Institut de Physique Nucl\'eaire ORSAY, Orsay, France}

\author {K.P. ~Adhikari} 
\affiliation{\ODU}
\author {D.~Adikaram} 
\affiliation{\ODU}
\author {M.~Aghasyan} 
\affiliation{\INFNFR}
\author {M.D.~Anderson} 
\affiliation{\GLASGOW}
\author {S. ~Anefalos~Pereira} 
\affiliation{\INFNFR}
% ...

%
\vskip 1.50in
\vskip 2in

\begin{abstract}
\centerline{\Large Abstract}
\vskip 0.2in

We report the first measurement of the photoproduction cross section of the $\phi$ meson in its neutral decay mode in the reaction $\gamma p \to p\phi(K_SK_L)$. The experiment was performed with a tagged photon beam of energy $1.6 \le E_\gamma \le 3.6$ GeV incident on a liquid hydrogen target of the CLAS spectrometer at the Thomas Jefferson National Accelerator Facility. The $p \phi$ final state is identified via reconstruction of $K_S$ in the invariant mass of two oppositely charged pions and by requiring the missing particle in the reaction $\gamma p \to p K_S X$ to be $K_L$. The presented results significantly enlarge the existing data on $\phi$-photoproduction. These data, combined with the data from the charged decay mode, will help to constrain different mechanisms of $\phi$ photoproduction.

\end{abstract}

\maketitle

\newpage

\section{Introduction}
\label{Introduction}
\hspace{0.5cm}

Strangeness production at low photon energy starting from threshold is very interesting for different reasons. On one hand it is an energy domain well above the region controlled by low-energy theorems~\cite{Klein}, and on the other hand it is far below the region of perturbative QCD~\cite{Stermn,Mueller,Farrar}. In this energy range various baryon resonances and coupled-channel effects may play an important role in strangeness photoproduction~\cite{Sag,Menze,Jude,Xie,An}. Moreover, it is important to conduct strangeness production experiments to establish the quantum numbers of the known states as well as to search for so-called missing baryon resonances ~\cite{Kiswandhi,Oh,Isgur,Jassen,Caps}.

Because the $\phi$ meson is a particle with hidden strangeness it can be produced without an associated hyperon. Also, due to the suppressed strange quark content of the nucleon, $\phi$-meson production through disconnected quark graphs with the OZI rule violation may be significantly enhanced by and therefore sensitive to processes involving gluons~\cite{Isgur1}. The same argument is also valid for $\phi$ production via decays of intermediate non-strange baryon resonances. Besides the sensitivity to the excitation of $s$-channel  baryon resonances, $\phi$ meson production is  an excellent channel to study subprocesses underlying the photoproduction of vector mesons in general.

The $\phi$ production mechanism depends on the four-momentum transfer $t=(P_{\gamma}-P_{\phi})^2$, where $P_{\gamma}$ and $P_{\phi}$ are the four-momenta of the incoming photon and the outgoing $\phi$ meson. At low $t$ the reaction $\gamma p \to \phi p$ can be diffractive and proceed with Pomeron exchange and suppressed $\pi$ and $\eta$ exchange, while the contribution of $\pi$ and $\eta$ exchange diagrams, as well as production of the $\phi$ via decay of excited intermediate baryon resonances, are more relevant at higher values of $t$ ~\cite{Titov03,Titov99,Titov07,Wlm}. However, no information on the coupling of baryon resonances to this channel is presently available~\cite{PDG}.

\begin{figure}[htb!]
    \includegraphics[width=3.6in]{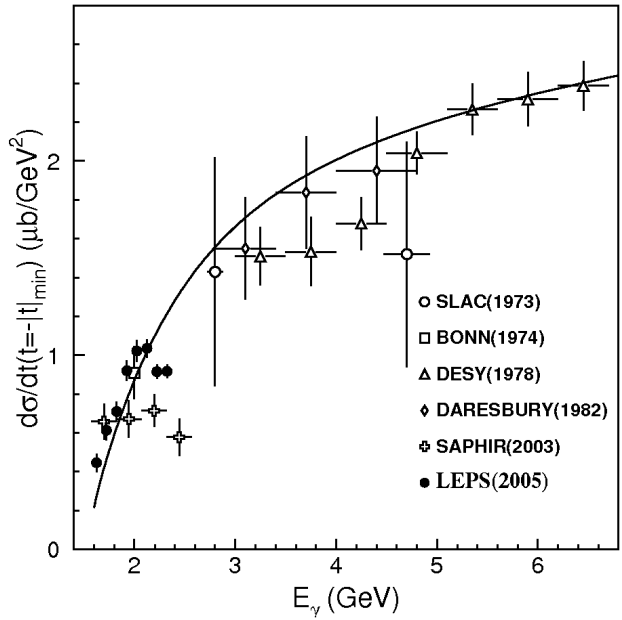}  
     \caption {Measured $\phi$ photoproduction cross section for the charged kaon decay channel, $(d\sigma/dt)$, extrapolated to $t=t_{min}$, as a function of incident photon energy $E_\gamma$ ~\cite{LEPS}. The solid line shows the cross section calculated according to the model of Titov and Lee~\cite{Titov03}.}
 \label{fig:world_data}
\end{figure}

Until now the $\phi$ photoproduction cross section has only been measured in a very limited kinematic range and only for the charged decay mode, $ \phi \to K^+K^−$ ~\cite{Saphir, LEPS, BONN, Daresbury, SLAC, DESY, Anciant, ZEUS}. The SAPHIR~\cite{Saphir} and LEPS~\cite{LEPS} data appear to show a peak in the cross section at about $1.8 \le E_\gamma\le 2.4$ GeV photon beam energy range as presented in Figure~\ref{fig:world_data}. In contrast, the calculation of Titov and Lee~\cite{Titov03} is smoothly increasing over this range.

Recently, Ryu {\it et al}~\cite{hosaka} argue that coupled-channel interactions may be responsible for the observed local enhancement of the forward cross section in a model dependent way. The main reason for this is the proximity of the energy threshold of two processes, $\gamma p\to K^+\Lambda(1520)$ and $\gamma p\to \phi p$. Therefore, the first intermediate process may affect the observed cross section of the second one due to the final state (or coupled-channel) interaction. 

An alternative explanation of the non-monotonic energy dependence of the forward cross section is proposed in Ref.~\cite{Kiswandhi}, where the authors interpret the existence of the bump in the differential cross sections at forward angles and near the reaction threshold as being due to an excitation of missing nucleon resonances with a non-negligible strangeness content.

Equally interestingly, at certain energies $\phi$ production and baryon production can also interfere. The $\gamma p \to p \phi \to pK \bar K$ reaction can interfere with the $\gamma p \to B K \to pK \bar K$ reaction (where $B$ is a baryon resonance that can decay to $p K(\bar K$)) since both have the same final state. For example, the final state $pK^-K^+$ can result from either $\phi$ photoproduction in the charged decay channel or from production of a $K^+$ and a prominent $\Lambda(1520)$ hyperon resonance with subsequent decay to $pK^-$. However, if  this is the case, then the photoproduction cross section associated with the p$K_SK_L$ final state in the neutral decay mode $\phi \to K_SK_L$ 
should differ from the charged decay mode as there is no similar prominent meson-baryon production mechanism in this case.

The analysis of $\phi$ photoproduction in the neutral decay channel is therefore a powerful tool to achieve a deeper understanding of the underlying photoproduction mechanisms. The high statistics CLAS data allows a measurement of the photoproduction  cross section of the $\phi$ meson in this decay mode over a wide kinematic range.

\section{Experiment}
\label{Experiment}
\hspace{0.5cm}

The present analysis is based on the g11a data set collected in 2004 using the CLAS detector at The Thomas Jefferson National Accelerator Facility (Jefferson Lab)~\cite{CLASDET}. The experiment was performed using a photon beam produced through bremsstrahlung from an electron beam of energy 4.02 GeV.
A scintillator hodoscope system combined with a dipole magnet was used to tag the photon time and energy in the range of 0.8 to 3.8 GeV with a resolution of 0.1$\%$ of the incident electron energy~\cite{Sober}. The photon timing resolution  was 120 ps. In this experiment the photon beam was incident on a 40 cm long liquid hydrogen target, centered 10 cm upstream from the center of the CLAS detector. The integrated luminosity  was approximately 80 pb$^{-1}$. 

A segmented scintillator detector (the Start Counter) surrounding the target was used for timing and triggering purposes~\cite{ST_counter}. 
Particles from the reaction were detected in the CLAS detector, consisting of six identical sectors, equipped with time-of-flight (TOF) scintillator counters~\cite{sdarticle2}, Electromagnetic Calorimeters~\cite{sdarticle}, Drift Chambers~\cite{sdarticle3,sdarticle4,sdarticle5,sdarticle6} and $\check{C}$erenkov Counters~\cite{sdarticle7}, covering nearly 4$\pi$ solid angle. The drift chambers consisted of three regions, located at three radial locations in each of the six sectors. The first region was placed before the strong magnetic field region, the second region was placed inside of a toroidal magnetic field used for momentum analysis, and the third region was placed after the magnetic field. The momentum resolution of the CLAS detector is momentum dependent and is of the order of $\Delta P/P$ $\sim 0.5\%$. Charged particle identification is based on simultaneous measurement of their momenta and time-of-flight.
 Due to various misalignments of the photon tagging system components, the measured photon energy had some inaccuracies causing reconstructed particle masses to deviate from their correct values. The momenta of detected particles also had inaccuracies because of the energy loss during their flight through the target and detectors. The discrepancies in the toroidal magnetic field map and in drift chamber survey information were causing some deviations in reconstructed particle momenta. To correct these inaccuracies, photon beam energy and charged particle momentum corrections were applied to data. The raw data used in this analysis were processed in the same way as in Ref.~\cite{DeVita}, including corrections for the energy loss of charged particles in the target, uncertainties in the magnetic field, and misalignments of the drift chambers.

\vspace{1.5cm}
\section{Data analysis}
\label{Data}

\subsection{Reconstruction of the final state}
\label{Evt}
\hspace{0.5cm}

In order to select the desired final state, we required that the events contained only three reconstructed charged particles in the final state: a proton, $\pi^+$ and $\pi^-$. These particles were identified by the standard CLAS event reconstruction software. The particle identification was based on the difference between the measured velocity $\beta_{meas}$ of the detected particle and the expected velocity $\beta_{th}$ calculated from the measured momentum and the masses of the different particles. The particle type was chosen based on the minimum difference between the measured $\beta_{meas}$ and $\beta_{th}$ velocities.

\begin{figure}[htb!]
     \includegraphics[width=3.6in]{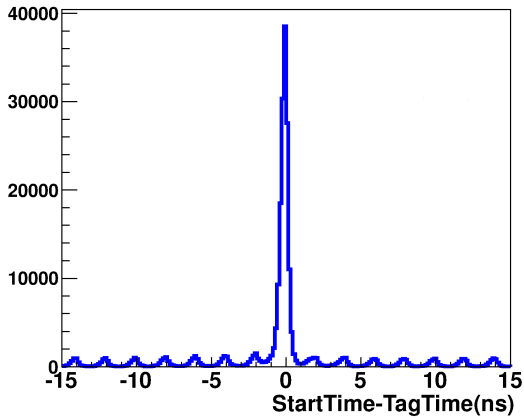}
     \caption
	{(Color Online) Difference between the time of the Start Counter hit closest to the event start time and the time of the tagger hits. A $\pm$2 ns cut was used to select good photons.}
\label{fig:st_time}
\end{figure}
 
The start time of the event at the interaction vertex was determined using the RF-signal from the accelerator injector that defined the timing of the electron bunches in the beam. The correct bunch containing the original electron that produced the interacting photon was selected by matching the timing information from the tagger counters and the CLAS Start Counter. Figure~\ref{fig:st_time} shows the distribution of the difference between the times from the best matching Start Counter hit and  the time of all hits in the photon tagger in the event. The $2$~ns structure corresponds to the bunch spacing of the electrons in the CEBAF beam that were designated for Hall B extraction. In order to suppress the accidental coincidences from the different beam bunches, we apply a cut on the time difference $|TagTime - StartTime|< 2$~ns. We also required that events contain only one photon detected in the tagger within this $2$~ns time interval. 

The momenta for protons and pions were required to be $P_{\pi} > 0.1$ GeV and $P_p > 0.35$ GeV, respectively, in order to avoid momentum values where charged track reconstruction becomes problematic. The minimum momentum cuts eliminate a very small fraction of pions (about 0.4$\%$) and protons (about 3$\%$) in our selected event sample.

$K_S$ mesons are reconstructed using the invariant mass $M(\pi^+\pi^-)$ of the $\pi^+ \pi^-$ system. Because the mean lifetime of the $K_L$ mesons is about 50 ns, few $K_L$ decayed in the CLAS detector. Thus, $K_L$ particles were reconstructed from the missing mass of the detected particles in the $\gamma p \rightarrow p \pi^+  \pi^- X$ reaction as $M_X^2=(P_{\gamma}+P_t-P_{\pi^+}-P_{\pi^-}-P_p)^2$, where $P_i$ are the four momenta of the photon, target proton, detected $\pi^+$, $\pi^-$ and proton, respectively. The missing mass of the $\gamma p \rightarrow p X$ system was used to select events coming from the decay of the $\phi$ meson.

To suppress the background under the $\phi$ meson, cuts were applied to the following distributions
\begin{itemize}
\item Invariant mass $M(\pi^+\pi^-)$ of the $\pi^+ \pi^-$ system
\item Missing mass $M_X$ of the $\gamma p \rightarrow p \pi^+  \pi^- X$ reaction
\end{itemize}

\begin{figure}[htb!]
    \includegraphics[width=3.0in]{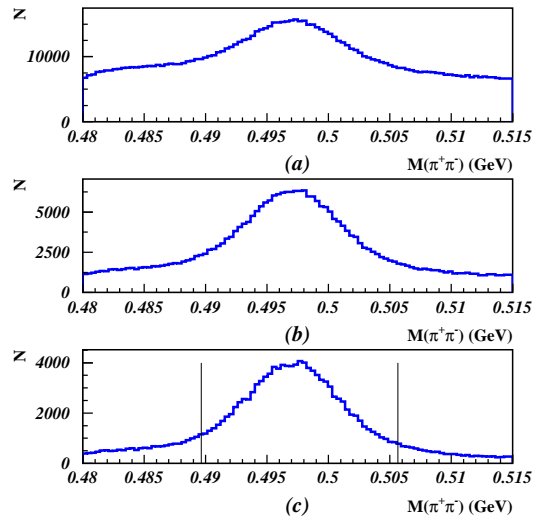}
     \caption
	{(Color Online) Invariant mass of the $\pi^+\pi^-$ system from the $\gamma p \rightarrow p \pi^+ \pi^-$ reaction: (a) with a cut on the mass of the missing $K_L$ from the $\gamma p \rightarrow p \pi^+ \pi^- X$ reaction $|M_X-M_{K_L}|<0.1$ GeV; (b) with a cut on the mass of the missing $K_L$ from the $\gamma p \rightarrow p \pi^+ \pi^- X$ reaction $|M_X-M_{K_L}|<0.015$ GeV; (c) with the previous cut and an additional cut on the missing mass of $\gamma p \rightarrow p X$, $|M_X - M_{\phi}|<0.02$ GeV.}
 \label{fig:kashort_clean}
\end{figure}

Figure~\ref{fig:kashort_clean} shows the invariant mass of the $\pi^+\pi^-$ system with wide and narrow cuts on $M_{K_L}$ and with a cut on $M_\phi$. $M(\pi^+\pi^-)$ is required to be within $2\sigma$ of the ${K_S}$ mass, $\vert M(\pi^+\pi^-) - M_{K_S} \vert \le 0.008$~GeV.

\begin{figure}[htb!]
    \includegraphics[width=3.0in]{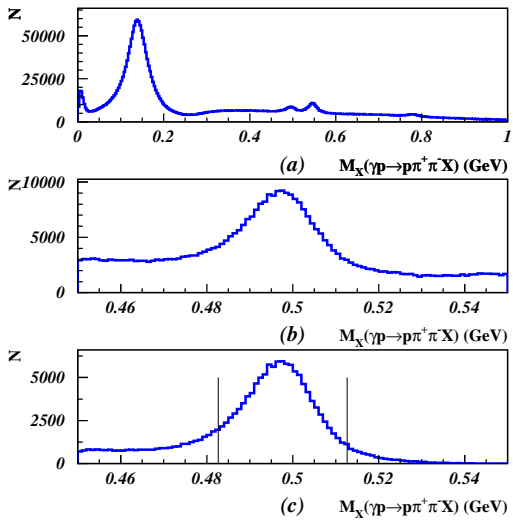}
     \caption
	{(Color Online) Missing mass of the $\gamma p \rightarrow p \pi^+ \pi^- X$ reaction: (a) with a cut on the invariant mass of the $\pi^+\pi^-$ system in the $\gamma p \rightarrow p \pi^+ \pi^-$ reaction $|M(\pi^+\pi^-)-M_{K_S}|<0.035$ GeV; (b) with a cut on the invariant mass of $\pi^+\pi^-$  $|M(\pi^+\pi^-)-M_{K_S}|<0.008$ GeV; (c) with the previous cut and an additional cut on the missing mass of $\gamma p \rightarrow p X$, $|M_X - M_{\phi}|<0.02$ GeV.}
 \label{fig:kalong_clean}
\end{figure}

Figure~\ref{fig:kalong_clean} shows the mass distribution of the missing $K_L$ in the $\gamma p \rightarrow \pi^+ \pi^- p X$ reaction with wide ($8\sigma$) and narrow ($2\sigma$) cuts on the $K_S$ mass and with a cut on $M_\phi$.
To select $K_L$ a cut was applied on the missing mass $M_{X}(\gamma p \rightarrow p \pi^+  \pi^- X)$ to be within $2\sigma$ of the $K_L$ mass: $|M_X-M_{K_L}|\leq 0.015$~GeV. 

\begin{figure}[htb!]
    \includegraphics[width=3.0in]{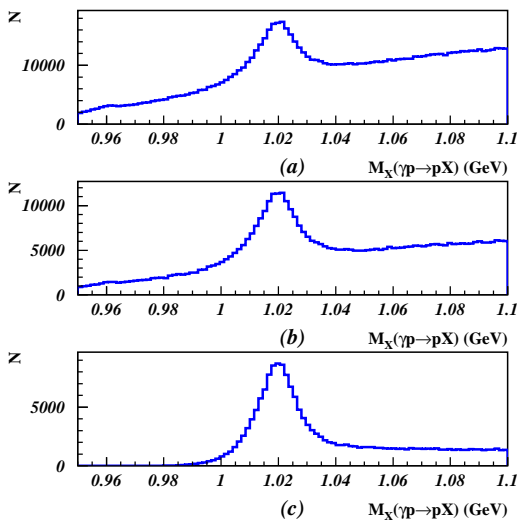}
     \caption {(Color Online) The missing mass of the $\gamma p \rightarrow p X$ reaction: (a) with cuts on the invariant mass of $\pi^+\pi^-$  $|M(\pi^+\pi^-)-M_{K_S}|<0.035$ GeV and on the mass of the missing $K_L$ from the $\gamma p \rightarrow p \pi^+ \pi^- X$ reaction $|M_X-M_{K_L}|<0.1$ GeV; (b) with a cut on the invariant mass of $\pi^+\pi^-$  $|M(\pi^+\pi^-)-M_{K_S}|<0.008$ GeV; (c) with the previous cut and an additional cut on the mass of the missing $K_L$ from the $\gamma p \rightarrow p \pi^+ \pi^- X$ reaction $|M_X-M_{K_L}|<0.015$ GeV.}
 \label{fig:phi_clean}
\end{figure}

Figure~\ref{fig:phi_clean} shows the distribution of the missing mass of the $\gamma p \rightarrow  p X$ reaction, where $\phi$ production is visible through the prominent peak at around $1.02$ GeV, after cuts: (a) $|M(\pi^+\pi^-)-M_{K_S}|<0.035$~GeV and $|M_X-M_{K_L}|<0.1$~GeV in the reaction $\gamma p \rightarrow p \pi^+ \pi^- X$, (b) $|M(\pi^+\pi^-)-M_{K_S}|<0.008$~GeV, (c) $|M(\pi^+\pi^-)-M_{K_S}|<0.008$~GeV and $|M_X-M_{K_L}|<0.015$~GeV. The signal to background ratio improves significantly after all three cuts are applied.

\subsection{Background subtraction}
\label{BG}
\hspace{0.5cm}

Figure~\ref{fig:phi_clean} shows that even after the application of the cuts described above there is still visible background left under the $\phi$ peak. The background is asymmetric, causing simple sideband subtraction techniques to be less reliable for signal--background separation. 

\begin{figure*}[h!]
    \includegraphics[width=7in]{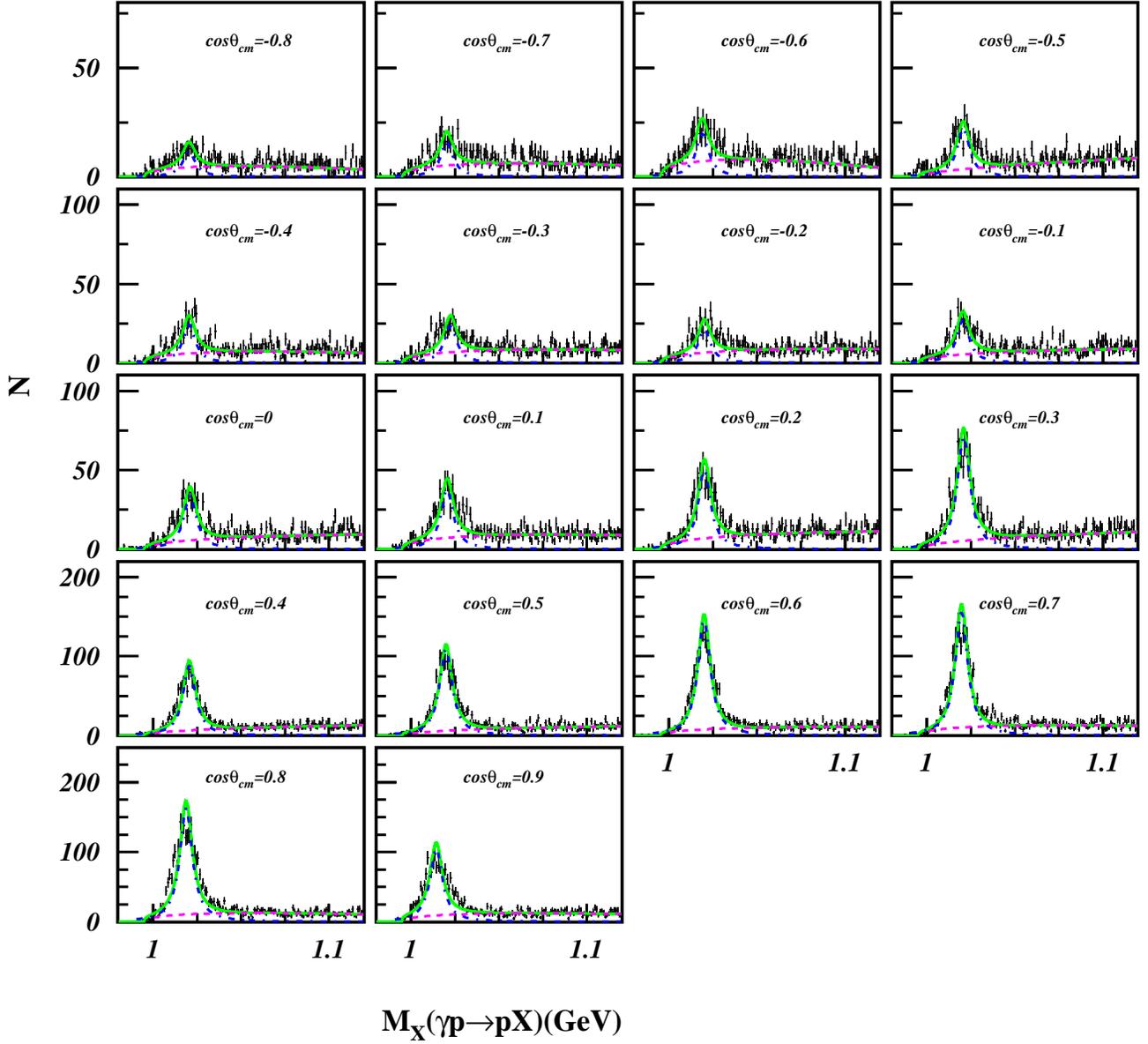}
     \caption
	{(Color Online) Signal and background separation for $1.8 \leq E_{\gamma} < 2.0$ GeV and different $\cos{\theta}_{cm}$ bins. The histograms (black) are the unweighted distributions from data. The dash-dotted line (blue) shows the signal from the fit. The dashed line (magenta) shows the background from the fit. The solid line (green) shows the overall fit of the mass distribution.}
\label{fig:fit1}
\end{figure*}

 Figure~\ref{fig:fit1} shows the fits of $M_X(\gamma p\rightarrow pX)$ mass distributions for a given photon energy range and in different $\cos{\theta}_{cm}$ bins, where $\theta_{cm}$ is the polar angle of the $\phi$ meson in the center-of-mass system of $\gamma p \rightarrow \phi p$. The black points are the experimental data shown with the statistical uncertainties. Voigt profile, which is a convolution of Lorentzian and Gaussian profiles, was used for the fit of the $\phi$ signal and to account for the broadening of the signal width due to the detector resolution effects

\begin{eqnarray}
S(m,E_{\gamma},\zeta) = {F(E_{\gamma},\zeta)} \frac{e^{-\frac{(m - \mu(E_{\gamma},\zeta))^2}{2\sigma^2(E_{\gamma},\zeta) }}}{\sigma(E_{\gamma},\zeta)} && \nonumber\\
\times \frac{\Gamma}{(m-\mu(E_{\gamma},\zeta))^2 + \Gamma^2}
\label{eq:fit_signal}
\end{eqnarray}

where $\Gamma$ is the width of the $\phi$, $\mu$ is the mean of the $\phi$ mass distribution and $\sigma$ is the standard deviation from the mean value for the particular $E_{\gamma}$ and $\cos{\theta}_{cm}$ bin, $m$ is the $\phi$ mass, $\zeta$=$\cos{\theta}_{cm}$ and $F$ is a fit parameter. The Gaussian part of the fit function accounts for the detector resolution impact on the signal width, and the Lorentzian profile describes the natural width of the $\phi$. The detector resolution for the mass was allowed to vary in the fit.

The background function used for the fit is of the following form: 
\begin{eqnarray}
B(m,E_{\gamma},\zeta)=a(E_{\gamma},\zeta)\sqrt{m^2 - 4m_K^2} \nonumber\\
+ b(E_{\gamma},\zeta)(m^2 - 4m_K^2),
\end{eqnarray}

\noindent where $a$ and $b$ are obtained from the fit to the data for each bin. Here $m_K=0.497648$ GeV is the mass of $K^0$ meson. 
The total statistics at very backward angles is marginal and extraction of the signal becomes unreliable. Therefore, these kinematic regions were disregarded.

The signal separation for the measurements is done on an event-by-event basis using the fit parameters of the $\phi$ mass distribution in different $E_\gamma$ and $\cos{\theta}_{cm}$ bins to weight each event with the signal probability coefficient defined as
\begin{equation}
  W=\frac{S(m,E_{\gamma},\zeta)}{S(m,E_{\gamma},\zeta) + B(m,E_{\gamma},\zeta)}.
\label{eq:bkg_weight}
\end{equation}

\subsection{Monte Carlo simulation}
\label{ACC}
\hspace{0.5cm}

In order to correct the experimental data for acceptance and efficiency, we developed an event generator and used it in conjunction with the simulation of the CLAS detector response. The Monte Carlo (MC) events are generated using the Titov and Lee theoretical model for $\phi$-photoproduction~\cite{Titov03}, which describes well the existing data for $1.6<E_{\gamma}<3.6$ GeV and for the range of $t = (P_\gamma - P_\phi)^2$ covered by this experiment. The $\phi$ then  decays as follows: $\phi \rightarrow K_S+K_L$ and $K_S \rightarrow \pi^+\pi^-$. The simulated final state of $p \pi^+ \pi^-$ is then passed through the GEANT Simulation program (GSIM) for the CLAS detector and is processed with the GSIM Post Processing (GPP) package, which takes into account missing detector channels and finite detector resolution. Finally, all the simulated events are reconstructed using the CLAS standard reconstruction code (RECSIS).

The $\phi$ decay angular dependencies  are taken from the Pomeron exchange model of Titov and Lee \cite{Titov03}. The $t$-dependence of the $\phi$ production is parametrized by the exponential form with $I\sim e^{-B(t-t_{min})}$, with a constant slope of $B=3.4$ GeV$^{-2}$.

The Monte Carlo events were generated using uniform photon energy distribution. The photon energy dependence was determined by fitting the reconstructed simulated events to the experimental yield.

The detector acceptance correction factor, here also referred to as acceptance correction, is obtained for every $E_\gamma$ and $\cos{\theta}_{cm}$ bin as the ratio of the number of reconstructed events to the number of simulated events. The relative uncertainty of the acceptance correction is estimated on a bin--by--bin basis by assuming a binomial distribution for the detection and reconstruction probability and using the following expression:
\begin{center}
\begin{equation}
{\sigma_{Acc}=\sqrt{\frac{(1-\frac{R}{N})}{\frac{R}{N}(N-1)}}},
\label{eq:variance}
\end{equation}
\end{center} 
where $N$ is the total number of simulated events, and $R$ is the number of accepted events. The average uncertainties due to the acceptance corrections are on the order of $\approx 5\%$ and are included into the statistical uncertainty of the measured values on a bin-by-bin basis.

\subsection{Fiducial cuts}
\label{FidCut} 
\hspace{0.5cm}

Both the experimental data and the detector acceptance were binned in two dimensions: the incident photon energy $E_{\gamma}$ and either $\cos{\theta}_{cm}$ or $|t - t_{min}|$.  The cross sections were averaged over the azimuthal angle. The bin sizes of the data and acceptance look-up tables were the same. In order to avoid the areas at the edges of the CLAS kinematic coverage, where our detector simulation did not reliably reproduce the experimental data, we developed a cut to eliminate those data bins. First, we found the $t$-bin with maximal value of acceptance for each energy bin. Next, we eliminated all $t$-bins  where the acceptance value was estimated to be less than $12\%$ from its maximal value for the same energy bin.

For this analysis the proton, $\pi^+$ and $\pi^-$ detection efficiencies were obtained using two reactions: $\gamma p \rightarrow p \pi^+ \pi^-$ and $\gamma p \rightarrow p \pi^+ \pi^- \pi^+ \pi^-$. Events were selected based on complete exclusivity, in case one particle was missed it was reconstructed by missing mass. The efficiencies were obtained as a function of the $\theta$ and $\phi$ angles, and momenta of the detected particles in the laboratory frame. We only used these efficiency values for cuts, and did not apply them to the data since the detector inefficiencies were already accounted for by using the GEANT-based detector simulation and GSIM Post Processing package (Section~\ref{ACC}). More details about the detection efficiency in this analysis can be found in~\cite{Thesis}.

The fiducial cuts were applied to eliminate regions of the detector where the particle detection and reconstruction efficiency changed rapidly and was less than 40$\%$. Additionally, events with kinematics corresponding to malfunctioning TOF scintillator paddles were excluded. These two cuts were applied both to data and MC reconstructed events.

\subsection{Normalization}
\label{Norm} 
\hspace{0.5cm}

The differential cross sections are calculated using
\begin{eqnarray}
\frac{d\sigma}{dt}  =  \left(\frac{A}{\mathcal{F}(E_{\gamma})\rho L N_A}\right)\frac{\mathcal{Y}(E_\gamma,t-t_{min})}{\Delta (t-t_{min})\eta(E_{\gamma},t-t_{min})}&& \nonumber\\
\times \frac{1}{BR(\phi \rightarrow  K_s K_L).}
\label{eqn:dif_cros_sect_tt}
\end{eqnarray}
Here $A$, $\rho$ and $L$ are the atomic weight, density and the length of the target, respectively. $N_A$ is Avogadro's number and $\mathcal{F}(E_{\gamma})$ is the total number of photons incident on the target in that photon energy bin. $\Delta (t-t_{min}) = 0.04$ GeV$^2$ is the $t-t_{min}$ bin size. $BR \approx 0.342$~\cite{PDG} is the branching ratio for the decay $\phi \rightarrow K_S K_L$. $\mathcal{Y}(E_{\gamma},t - t_{min})$ is the number of events in the given photon energy and $t-t_{min}$ bin, that passed all cuts (see Table~\ref{tab: kinematic_cuts}) after the background separation (see Section \ref{BG}). $\eta(E_{\gamma},t-t_{min})$ is the acceptance in the $(E_{\gamma},t-t_{min})$ bin. The cross sections $d\sigma/d\cos\theta_{cm}$ was obtained in the same way as a function of $\cos{\theta}_{cm}$ using the corresponding yields and acceptance correction factors and the $\cos{\theta}_{cm}$ bin size of $0.1$.

\begin{table}
\begin{center}
\caption{Table of cuts applied for data selection.}
\label{tab: kinematic_cuts}
\begin{tabular}{|l|l|}
\hline \hline
{\bf Cuts} & {\bf Description} \\ \hline
 Momentum cuts &  $P_{\pi^+,\pi^-}>0.1$ GeV, $P_p>0.35$ GeV\\ 
 $K_S$ selection & $|M(\pi^+\pi^-)-0.49765|\le 0.008$ GeV\\ 
 $K_L$ selection & $|M_X(p \pi^+ \pi^-)- 0.49765|\le 0.015$ GeV\\ 
 Efficiency cut & $\mathcal{E}_{p,\pi^+,\pi^-}>40\%$ \\
 TOF paddles & Malfunctioning TOF paddles \\ 
 Fiducial cuts & Sec.~\ref{FidCut}\\
 Timing cut & $|TAG_{time}-ST_{time}|< 2$ ns \\ 
 Acceptance cuts& $\eta/\eta_{max}>12\%$\\\hline \hline
\end{tabular}
\end{center}
\end{table}

The photon flux for the experiment was obtained using the CLAS  {\it gflux} software package ~\cite{BallPasyuk}. In the g11a experiment a linear dependence of the normalized photon yield on the beam current was observed. This was suggested to be due to the malfunctioning of the hardware used to estimate the Data Acquisition (DAQ) live time. A correction factor was obtained fitting the yield dependence on the beam current with a line and correcting for the drop in the normalized yield. This factor was 1.187 at the electron beam current of 65 nA~\cite{DeVitaMarcoFlux}.

As mentioned in Section ~\ref{Evt}, it was required to have only one photon present in the tagger within the $\pm$2 ns time interval between the tagger and the Start Counter: $|TAG_{time}-ST_{time}|< 2$ ns. This cut removed some good events from the data. This loss of events was corrected for by normalizing the overall yield with two correction factors: {\it multiple hits} and {\it time window}. The {\it multiple hits} correction was found to be on the order of $18\%$ for our process and was obtained by comparison of the $\phi$ meson yields with and without the requirement to have only one good photon detected by the tagger for the event. The $6\%$ {\it time window} correction is obtained by comparing the yields with $\Delta t=2$ ns cut and with $\Delta t=15$ ns (no cut)~\cite{DeVitaMarcoFlux}.

\subsection{Systematic uncertainties}
\label{SYST}
\hspace{0.5cm}

 The summary of the estimated relative systematic uncertainties of the measured cross sections is given in Table~\ref{tab:syst_errors_tab}.

As described in Section \ref{BG}, the $\phi$ meson background was estimated from the fits of the $\phi$ mass distributions in each photon energy and $\cos{\theta}_{cm}$ bin. The systematic uncertainty due to background subtraction was estimated from the uncertainties of the fit parameters and in average was on the order of 10\%.
 
 To eliminate low efficiency regions of the detector a cut was applied on the single-particle detection efficiencies.  This cut was applied to both the reconstructed Monte Carlo events and the data (see Section \ref{FidCut}). The systematic uncertainty of this cut was estimated by measuring the acceptance-corrected cross sections for three different efficiency cuts: $30\%$, $35\%$ and $40\%$. The systematic uncertainty was estimated bin--by--bin as the standard deviation of the three resulting cross sections.  The average systematic uncertainty due to the efficiency cut was about $6\%$. 

In the previous section we described a correction factor due to the beam current dependence of the normalized yield for the g11a experiment. This factor was obtained by a fit and has an uncertainty of about $3\%$.

\begin{table}[!ht]
\begin{center}
\caption{Summary of systematic uncertainties.}
\label{tab:syst_errors_tab}
\begin{tabular}{ |l |c | }
\hline
\hline
   {\bf Uncertainty source} & {\bf Uncertainty} \\ \hline
   Background Subtraction & 10$\%$ \\ 
   Photon Normalization & 7.7$\%$~\cite{Williams2}\\
   Particle Detection Efficiency Cut & 2--6$\%$ \\
   Current--Dependent Correction &  3$\%$~\cite{Williams} \\
   $\phi \rightarrow K_L K_S$ Branching Fraction & 0.6$\%$~\cite{PDG}\\
   Photon Transmission Efficiency &  0.5$\%$~\cite{Williams2} \\
   Target Length & 0.13$\%$ \\
   Target Density &  0.11$\%$~\cite{Williams} \\  
\hline
\hline
\end{tabular}
\end{center}
\end{table}

 The photon flux normalization uncertainty of $7.7\%$ was obtained from comparison of the cross sections for a set of reference reactions ($K^+\Lambda$, $p\omega$, $p\eta$) measured using the current (g11a) data set and other experimental data~\cite{Williams2,McCracken,Williams3}.

 The overall systematic uncertainty of the differential cross section averaged over all bins is about  $14\%$.

\section{Experimental results}
\label{CSC}
\hspace{0.5cm}

The differential cross section for $\phi$ meson photoproduction was measured as a function of $|t-t_{min}|$ and $\cos{\theta}_{cm}$ for different photon energy bins. The differential cross section $d\sigma/dt$ was also measured as a function of $E_{\gamma}$ for different $\cos{\theta}_{cm}$ bins. 

\subsection{The $t$-dependence}

The differential cross sections $d\sigma/dt$ were obtained for different four--momentum transfer $t-t_{min}$ bins and for 0.1 GeV photon energy bins in the photon energy range 1.6--3.6 GeV. 

\begin{figure*}[htb!]
  \includegraphics{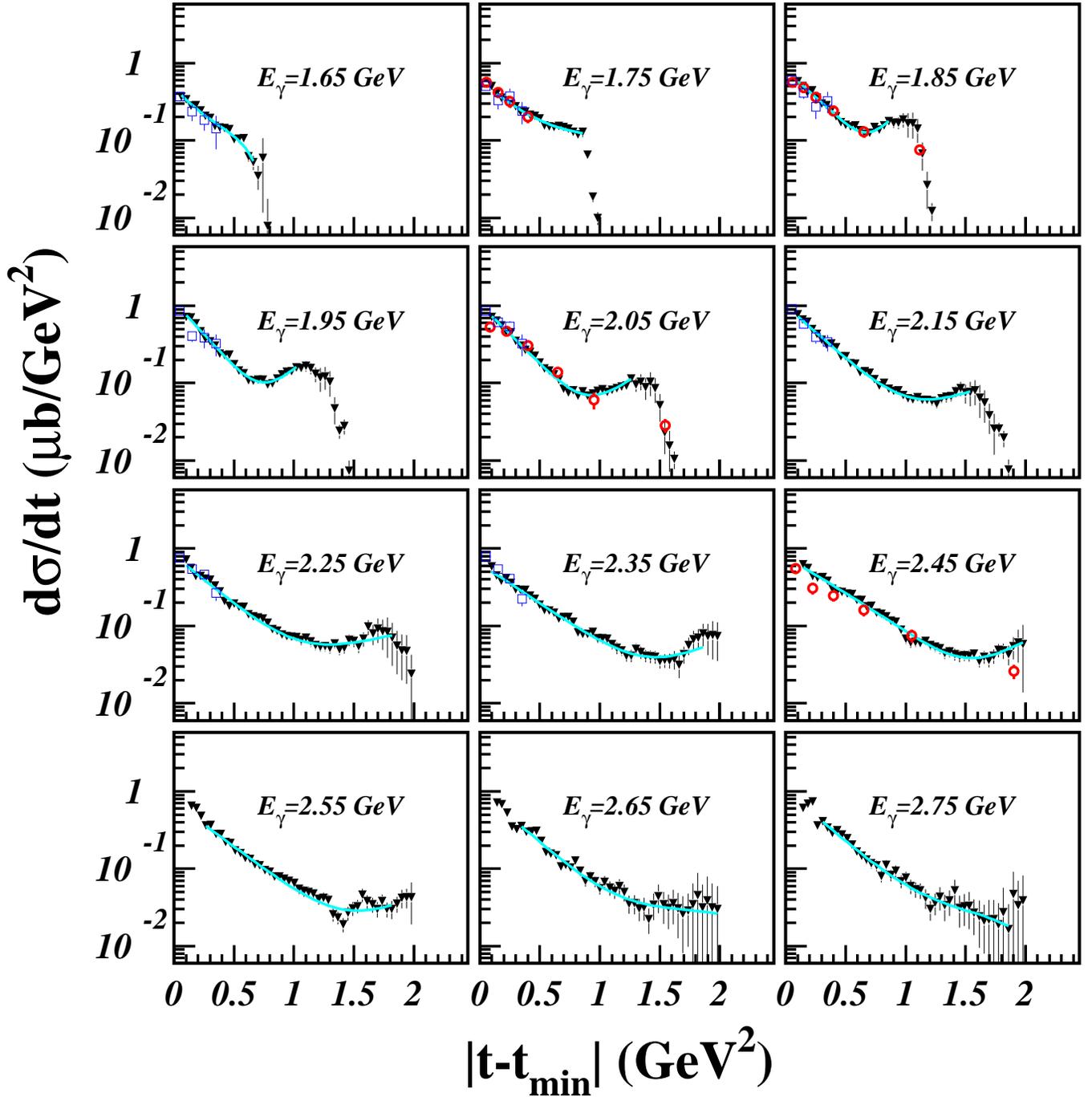}
     \caption
	{(Color Online)  Differential cross section $d\sigma/dt$ plotted versus $|t-t_{min}|$ for different photon energy bins in the $1.6 \leq E_{\gamma} < 2.8$ GeV range. The filled downward triangles (black) are the results of the current CLAS(2013) analysis for the neutral decay mode of the $\phi$.  The open circles (red) are the SAPHIR~\cite{Saphir} data for the charged mode topology. The open squares (blue) are the LEPS~\cite{LEPS} data for the charged mode topology. The solid lines (cyan) show the fit function.}
\label{fig:dat_mc1}
\end{figure*}

\begin{figure*}[htb!]
 \includegraphics{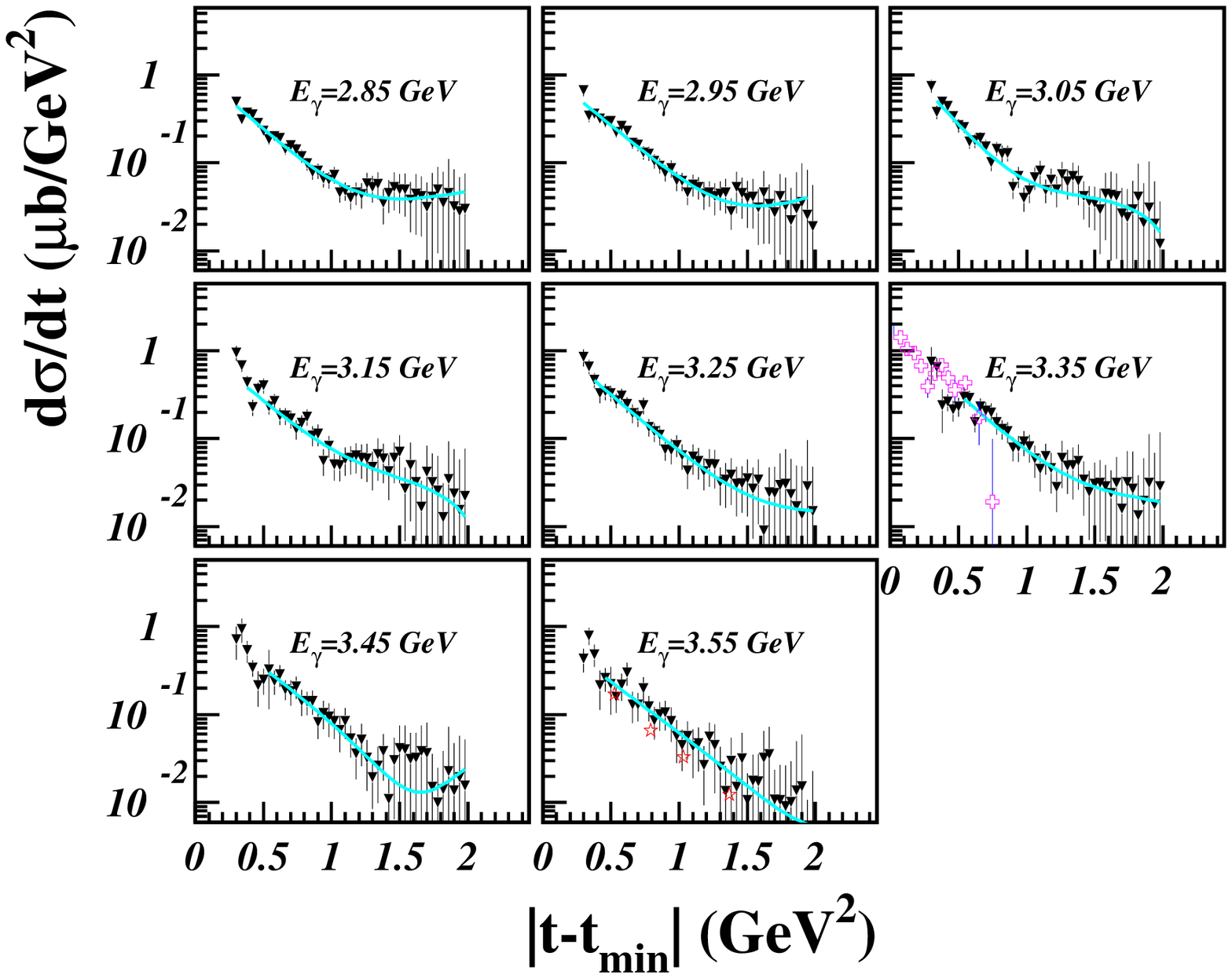}
\vskip -1.5in
     \caption
	{(Color Online) Differential cross section $d\sigma/dt$ plotted versus $|t-t_{min}|$ for different photon energy bins in the $2.8 \leq E_{\gamma} \leq 3.6$ GeV range. The filled triangles (black) are the results of the current CLAS(2013) analysis for the neutral decay mode of the $\phi$. The open crosses (magenta) are the DARESBURY~\cite{Daresbury} data for the charged mode. The open asterisks (red) represent the CLAS(2000)~\cite{Anciant} results for charged decay mode of the $\phi$ meson. The solid lines (cyan) show the fit function.}
\label{fig:dat_mc2}
\end{figure*}

Figures~\ref{fig:dat_mc1} and~\ref{fig:dat_mc2} show the $|t-t_{min}|$ dependences of the differential cross section for different photon energy bins. The world data for the charged decay channel are also included for comparison. The neutral decay mode data are fitted with an exponential function (solid lines) to obtain the differential cross section at $t=t_{min}$ and the slope of the $t$-dependence of $\phi$ photoproduction. These measurements show that in addition to the fast exponential  fall-off, which is  characteristic of $t$-channel exchange, at the higher values of the photon beam energy (about $2$ GeV) the cross section distribution starts flattening at larger values of $t$. This is indicative of the possible presence of other mechanisms of $\phi$ production, such as excitation and decay of intermediate nucleon resonances.

\begin{figure*}[htb!]
   \includegraphics[width=7in]{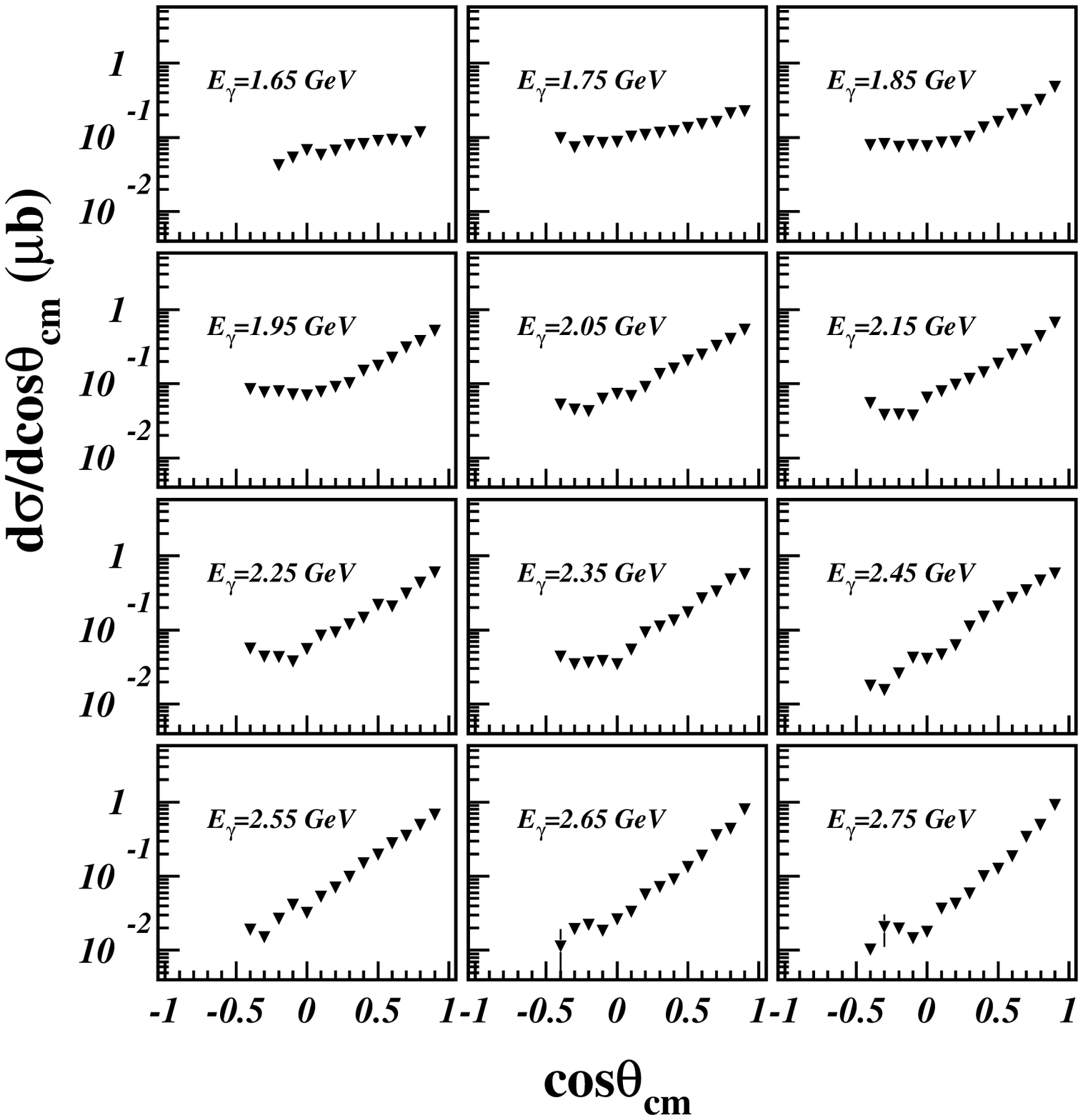}
     \caption{The $\cos{\theta}_{cm}$ dependence of the $\phi$ meson differential cross section $d\sigma/d\cos\theta_{cm}$ for different photon beam energy bins $1.6\leq E_{\gamma}<2.8$ GeV. The $\cos{\theta}_{cm}$ bin size is 0.1.}
 \label{fig:cosnk1}
\end{figure*}

\begin{figure*}[htb!]
  \includegraphics[width=7in]{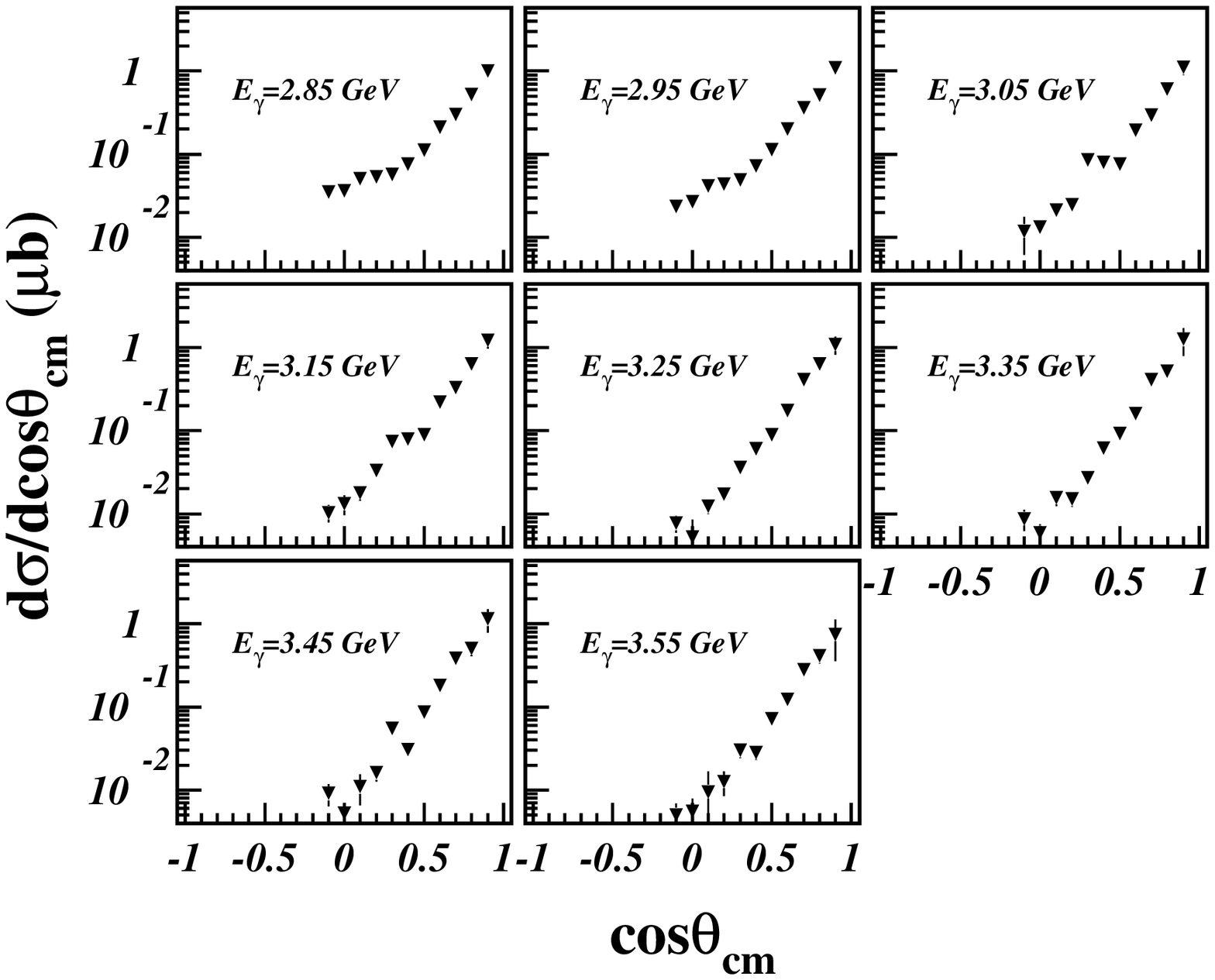}
  \vskip -1.5in
     \caption{The $\cos{\theta}_{cm}$ dependence of the $\phi$ meson differential cross section $d\sigma/d\cos\theta_{cm}$ for different photon beam energy bins $2.8 \leq E_{\gamma} \leq 3.6$ GeV. The $\cos{\theta}_{cm}$ bin size is 0.1.}
\label{fig:cosnk2}
\end{figure*}

Figures~\ref{fig:cosnk1}--\ref{fig:cosnk2} show the differential cross section $d\sigma/d\cos\theta_{cm}$ plotted as a function of $\cos{\theta}_{cm}$. The results are plotted for 0.1 GeV photon energy bins.

\hspace{0.5cm}

\subsection{The $E_{\gamma}$-dependence}
For this work, the differential cross section $d\sigma/dt$ for the process will be given as
\begin{equation}
\frac{d\sigma}{dt} = \frac{1}{2}\left(\frac{1}{E_{\gamma}|\overrightarrow{p}_{\phi}|}\right)_{cm}\left(\frac{d\sigma}{d\cos\theta_{cm}}\right).
\label{eq:diff_csc_conv}
\end{equation}

\begin{figure*}[htb!]
    \includegraphics{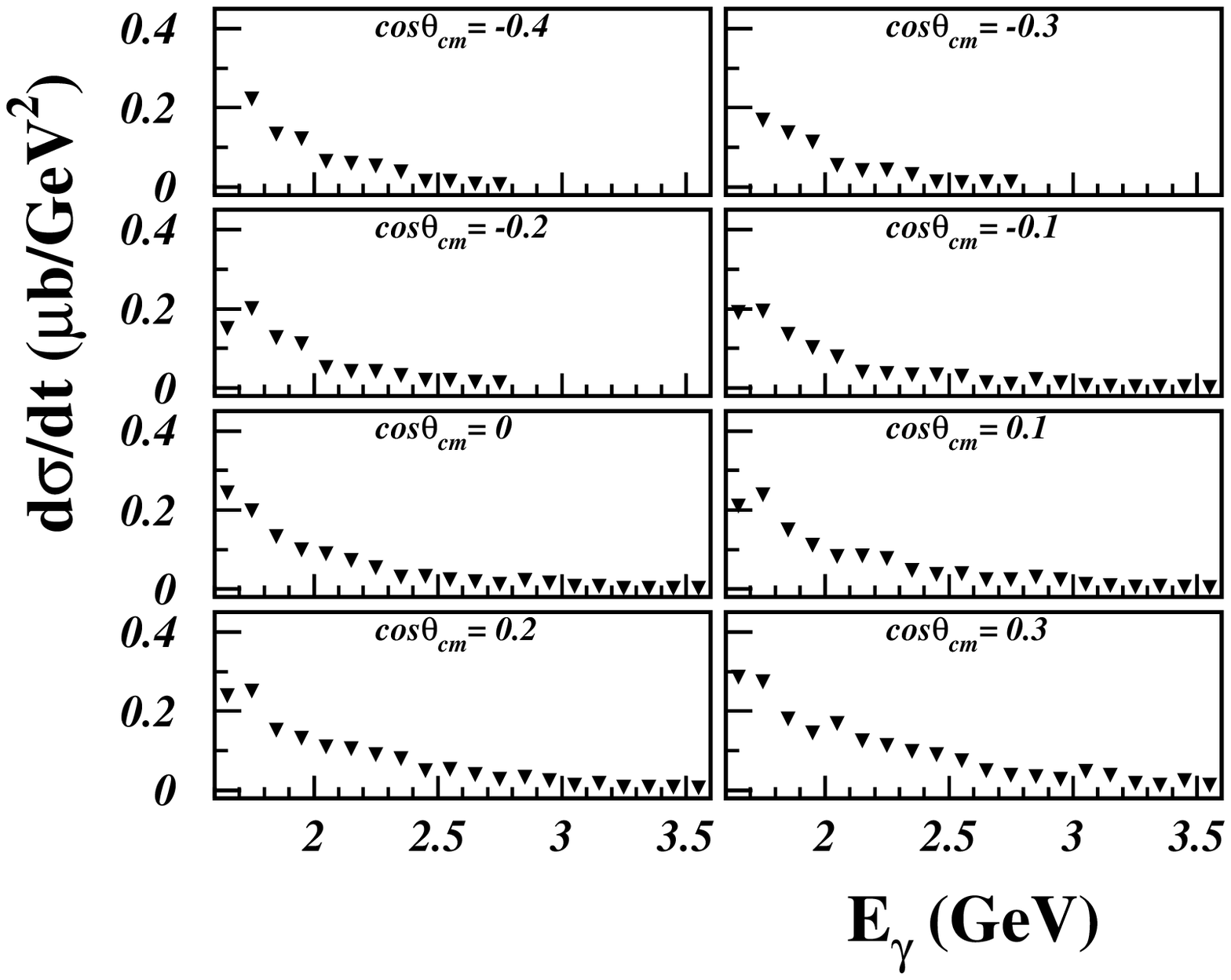}
     \caption
	{The $E_{\gamma}$ dependence of the $\phi$ meson differential cross section $d\sigma/dt$ in different $\cos{\theta}_{cm}$ bins for $-0.45 \leq \cos{\theta}_{cm} < 0.35$. }
 \label{fig:wnkar1}
\end{figure*}

\begin{figure*}[htb!]
     \includegraphics{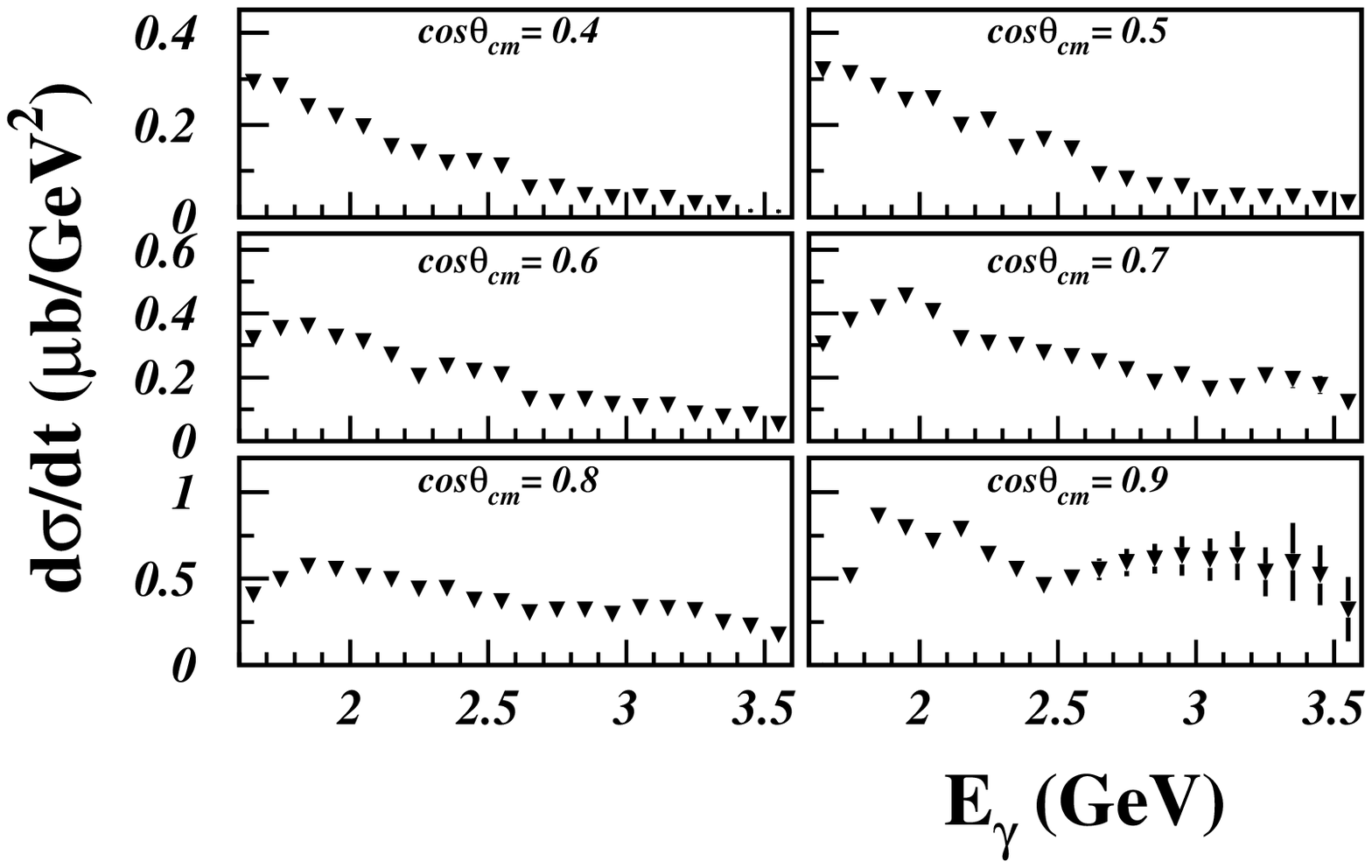}
\caption
	{The $E_{\gamma}$ dependence of the $\phi$ meson differential cross section $d\sigma/dt$ in different  $\cos{\theta}_{cm}$ bins for $0.35 \leq \cos\theta_{cm} < 0.95$.}
 \label{fig:wnkar2}
\end{figure*}

Figures~\ref{fig:wnkar1}--\ref{fig:wnkar2} show the $E_{\gamma}$ dependence of the differential cross section $d\sigma/dt$ in different $\cos{\theta}_{cm}$ bins. First, $d\sigma/dt$ is measured at different photon energies $E_{\gamma}$ (for 0.05 GeV wide bins of $E_{\gamma}$).

In forward angle $\cos{\theta}_{cm}$ bins, corresponding to the low-$t$ region, there appears to be a local enhancement in $d\sigma/dt$ in the range of $2.0<E_{\gamma}<2.4$~GeV and a maximum at about $E_{\gamma}\approx2.2$ GeV. In this angular region, at higher energies ($E_{\gamma}>2.6$ GeV) the cross section behavior is almost constant, as expected by Pomeron exchange mechanism dominance in $\phi$ production~\cite{Titov03}.

\subsection{Total cross section} 
\label{TOTAL}
\begin{figure*}[htb!]
    \includegraphics[width=3.5in]{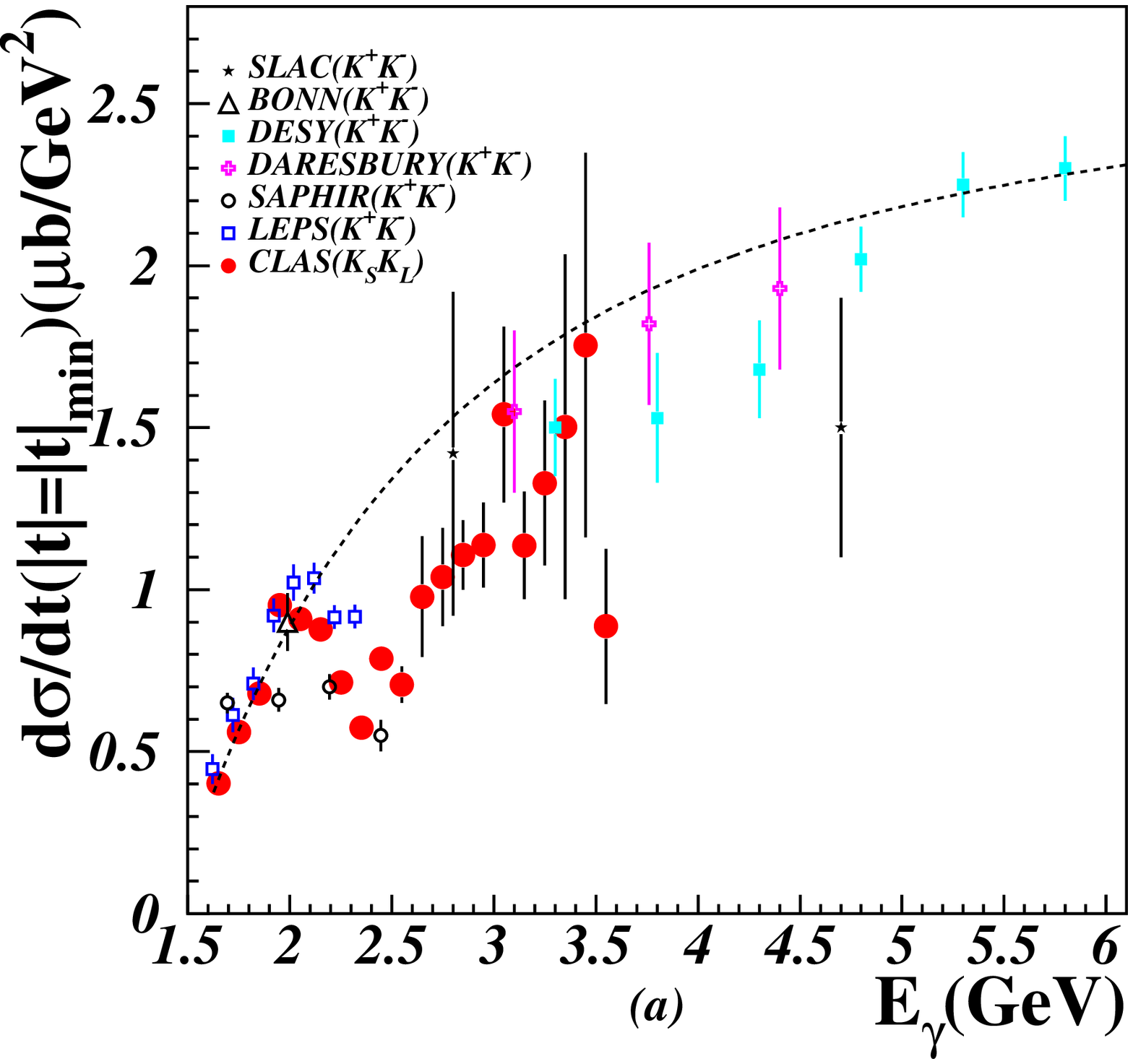}
    \includegraphics[width=3.5in]{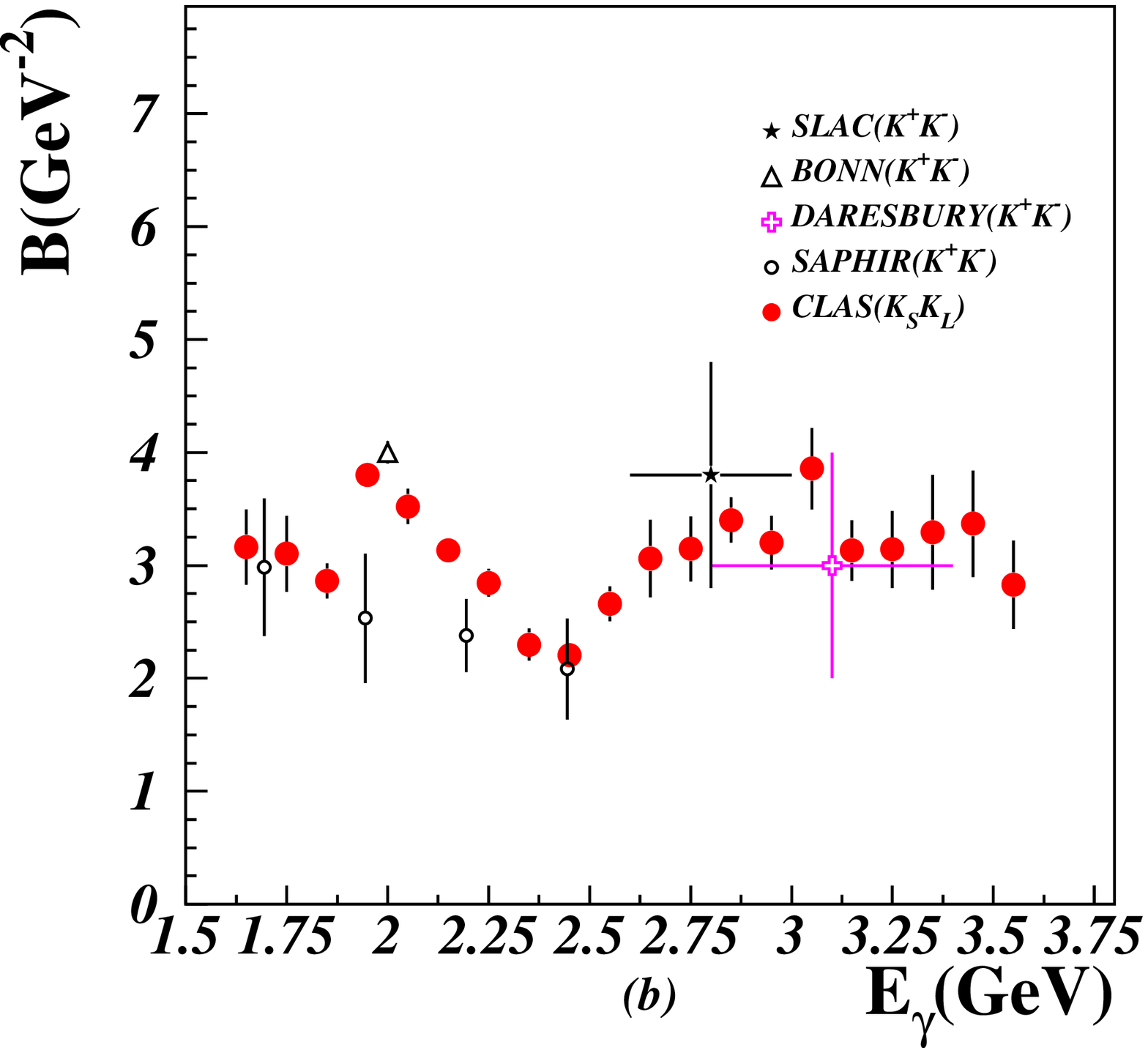}
     \caption
	{(Color Online) (a) Differential cross section $d\sigma/dt$ at $t=t_{min}$ and (b) $\phi$ meson photoproduction slope $B$ for the neutral decay mode plotted as a function of the photon beam energy $E_{\gamma}$. The error bars represent the statistical uncertainties. The filled circles (red) are the current CLAS (2013) data in the neutral mode. For the charged decay mode, the open circles (black) are the SAPHIR~\cite{Saphir} results, the open squares (blue) are the LEPS~\cite{LEPS} results,  the open crosses (magenta) are the DARESBURY~\cite{Daresbury} data, the asterisks (black) are the SLAC~\cite{SLAC} data, the open upward triangles (black) are the BONN~\cite{BONN} data, and the filled squares (cyan) are the DESY~\cite{DESY} data. The dashed curve represents the prediction of a model including the Pomeron trajectory, $\pi$ and $\eta$ exchange processes~\cite{Titov03}.}
 \label{fig:csc_results}
\end{figure*}

To determine the total cross section, an exponential function of the form $Ae^{-B|t-t_{min}|} + P(3)$, where $P(3)$ is a third order polynomial function, is used to fit the $|t-t_{min}|$ dependence of the differential cross section $d\sigma/dt$. The fit range was selected to have the $\chi^2/ndf$ value closest to $1$. The extracted values of the differential cross section at $t=t_{min}$ and the $t$ slope from the fit depend on the range of the fit and vary within $10\%$. The higher $t$ regions, where the cross section increases and deviates from the exponential behavior, were not included in the fit. Figures~\ref{fig:csc_results}(a) and (b) show the total cross section and the $B$ slopes for the different photon energy bins, respectively. The total cross section still has some enhancement in the photon beam energy range of about 1.8--2.3 GeV just as observed in the charged--mode decay. The photon beam energy dependence of the slope of the $t$ distribution also has some local structure in the photon beam energy range of 1.8--2.3 GeV.

\section{Conclusions}
\label{Conclude}
\hspace{0.5cm}

Differential cross sections for photoproduction of the $\phi$ meson via its neutral decay channel in the reaction $\gamma p\to p \phi(K_{S}K_L)$ have been measured for the first time. The $t$-dependence of the cross section at fixed photon energy indicates that the mechanism of $\phi$ production deviates from an exponential behavior at higher values of $t$. Detailed model calculations will be required to estimate the relative contribution to the $\phi$ production stemming from Pomeron, $\pi$ and  $\eta$ exchanges. The presented data will help to constrain the contribution of intermediate $s$-channel nucleon resonances and final state interactions.

On the other hand, the energy dependence of the differential cross section at the forward limit shows a similar local maximum as was observed previously in photoproduction of the $\phi$ meson in the charged decay mode.  There are different explanations for this local maximum based on either coupled-channel effects, excitation of intermediate nucleon resonances or  interference with $\Lambda(1520)$ production as was proposed in Refs.~\cite{Kiswandhi,hosaka,Titov99,Titov07}.

A full understanding of the mechanism of $\phi$ photoproduction at low energy and especially the energy 
dependence of the forward cross sections will require systematic comparison of theoretical model predictions 
with experimental data. Theoretical models for this purpose should include not only the $\phi$ photoproduction 
in both decay modes, but also the photoproduction cross sections of associated $pK(\bar K)$ system as 
meson-baryon final states are significantly different in these two decay modes.

\section{Acknowledgements}

We would like to acknowledge the outstanding efforts of the staff of the Accelerator and
the Physics Divisions at Jefferson Lab that made the experiment possible.

This work was supported in part by the Italian Istituto Nazionale di Fisica Nucleare, the French Centre
National de la Recherche Scientifique and Commissariat \`a l'Energie Atomique, the United Kingdom's Science and Technology Facilities Council (STFC), the U.S. Department of Energy and National Science Foundation, and the Korea Science and Engineering
Foundation. The Southeastern Universities Research Association (SURA) operates the Thomas
Jefferson National Accelerator Facility for the United States Department of Energy under
contract DEAC05-84ER40150.

\newpage
\clearpage
\onecolumngrid
\appendix
\section{}
\label{sec:Difcscval}
\setcounter{table}{0}
\renewcommand\thetable{\Alph{section}.\arabic{table}}
\begin{center}
\begin{table*}[h!]
\captionsetup{width=17cm} 
	\caption{$d\sigma/dt$ $(\mu b/GeV^2)$ vs. $|t-t_{min}|$ $(GeV^2)$ for different photon beam energies. Superscripts are statistical errors and subscripts are systematic errors.}
	\begin{tabular}{|c|c|c|c|c|c|c|}
	\hline \hline
	$E_{\gamma}$ $(GeV)$ & $\sqrt{s}$ $(GeV)$ & \multicolumn{5}{|c|}{$|t-t_{min}|$ $(GeV^2)$} \\
	\cline{3-7}
	 & &  $0.060$ & $0.100$ & $0.140$ & $0.180$ & $0.220$ \\
	\hline
      $1.6500$ & $1.9942$ & $0.3367^{\pm0.0168}_{\pm0.0677}$ & $0.2808^{\pm0.0133}_{\pm0.0385}$ & $0.2853^{\pm0.0135}_{\pm0.0103}$ & $0.2411^{\pm0.0119}_{\pm0.0173}$ & $0.2059^{\pm0.0108}_{\pm0.0104}$ \\
      $1.7500$ & $2.0407$ & $0.5033^{\pm0.0210}_{\pm0.0331}$ & $0.4052^{\pm0.0154}_{\pm0.0385}$ & $0.3562^{\pm0.0138}_{\pm0.0298}$ & $0.3262^{\pm0.0130}_{\pm0.0189}$ & $0.2929^{\pm0.0115}_{\pm0.0218}$ \\
      $1.8500$ & $2.0861$ & $0.5815^{\pm0.0252}_{\pm0.0682}$ & $0.4991^{\pm0.0196}_{\pm0.0443}$ & $0.4159^{\pm0.0156}_{\pm0.0495}$ & $0.4453^{\pm0.0169}_{\pm0.0701}$ & $0.3571^{\pm0.0145}_{\pm0.0240}$ \\
      $1.9500$ & $2.1306$ & $0.7171^{\pm0.0361}_{\pm0.1199}$ & $0.7058^{\pm0.0292}_{\pm0.0266}$ & $0.5945^{\pm0.0240}_{\pm0.0471}$ & $0.4673^{\pm0.0195}_{\pm0.0210}$ & $0.3744^{\pm0.0162}_{\pm0.0442}$ \\
      $2.0500$ & $2.1742$ & $0.7187^{\pm0.0486}_{\pm0.1023}$ & $0.6419^{\pm0.0311}_{\pm0.0344}$ & $0.5528^{\pm0.0255}_{\pm0.0745}$ & $0.4624^{\pm0.0205}_{\pm0.0351}$ & $0.4505^{\pm0.0204}_{\pm0.0307}$ \\
      $2.1500$ & $2.2170$ & $0.7754^{\pm0.0710}_{\pm0.0604}$ & $0.6640^{\pm0.0415}_{\pm0.0261}$ & $0.6215^{\pm0.0331}_{\pm0.0426}$ & $0.5022^{\pm0.0230}_{\pm0.0248}$ & $0.4302^{\pm0.0212}_{\pm0.0231}$ \\
      $2.2500$ & $2.2589$ & $0.7271^{\pm0.0846}_{\pm0.1039}$ & $0.5615^{\pm0.0367}_{\pm0.0389}$ & $0.4425^{\pm0.0248}_{\pm0.0408}$ & $0.4399^{\pm0.0269}_{\pm0.0380}$ & $0.4093^{\pm0.0238}_{\pm0.0456}$ \\
      $2.3500$ & $2.3001$ & $0.5836^{\pm0.0871}_{\pm0.0834}$ & $0.4521^{\pm0.0379}_{\pm0.0204}$ & $0.3822^{\pm0.0275}_{\pm0.0366}$ & $0.4159^{\pm0.0293}_{\pm0.0300}$ & $0.3583^{\pm0.0273}_{\pm0.0192}$ \\
      $2.4500$ & $2.3405$ & - & - & - & $0.6228^{\pm0.0467}_{\pm0.0316}$ & $0.5617^{\pm0.0400}_{\pm0.0682}$ \\
      $2.5500$ & $2.3803$ & - & - & - & $0.6534^{\pm0.0536}_{\pm0.0519}$ & $0.6069^{\pm0.0504}_{\pm0.1900}$ \\
      $2.6500$ & $2.4193$ & - & - & - & $0.7104^{\pm0.0722}_{\pm0.0443}$ & $0.6868^{\pm0.0659}_{\pm0.0301}$ \\
      $2.7500$ & $2.4578$ & - & - & - & $0.6062^{\pm0.0753}_{\pm0.0371}$ & $0.6965^{\pm0.0733}_{\pm0.0332}$ \\
\hline \hline
\end{tabular}
\label{table:table1}
\end{table*}
\end{center}

\begin{center}
\begin{table*}[h!]
	\caption{$d\sigma/dt$ $(\mu b/GeV^2)$ vs. $|t-t_{min}|$ $(GeV^2)$ for different photon beam energies. Superscripts are statistical errors and subscripts are systematic errors.}
	\begin{tabular}{|c|c|c|c|c|c|c|}
	\hline\hline
	$E_{\gamma}$ $(GeV)$ & $\sqrt{s}$ $(GeV)$ & \multicolumn{5}{|c|}{$|t-t_{min}|$ $(GeV^2)$} \\
	\cline{3-7}
	 & & $0.260$ & $0.300$ & $0.340$ & $0.380$ & $0.420$ \\
	\hline
      $1.6500$ & $1.9942$ & $0.1939^{\pm0.0107}_{\pm0.0251}$ & $0.1536^{\pm0.0092}_{\pm0.0081}$ & $0.1456^{\pm0.0092}_{\pm0.0076}$ & $0.1441^{\pm0.0101}_{\pm0.0214}$ & $0.1393^{\pm0.0110}_{\pm0.0214}$ \\
      $1.7500$ & $2.0407$ & $0.2734^{\pm0.0111}_{\pm0.0256}$ & $0.2335^{\pm0.0102}_{\pm0.0099}$ & $0.2345^{\pm0.0113}_{\pm0.0166}$ & $0.2133^{\pm0.0100}_{\pm0.0248}$ & $0.2082^{\pm0.0103}_{\pm0.0124}$ \\
      $1.8500$ & $2.0861$ & $0.2992^{\pm0.0128}_{\pm0.0438}$ & $0.2842^{\pm0.0122}_{\pm0.0130}$ & $0.2484^{\pm0.0114}_{\pm0.0186}$ & $0.2034^{\pm0.0096}_{\pm0.0236}$ & $0.1707^{\pm0.0087}_{\pm0.0306}$ \\
      $1.9500$ & $2.1306$ & $0.3481^{\pm0.0159}_{\pm0.0344}$ & $0.3052^{\pm0.0140}_{\pm0.0103}$ & $0.2398^{\pm0.0116}_{\pm0.0184}$ & $0.2287^{\pm0.0119}_{\pm0.0111}$ & $0.2333^{\pm0.0127}_{\pm0.0266}$ \\
      $2.0500$ & $2.1742$ & $0.3503^{\pm0.0177}_{\pm0.0428}$ & $0.3038^{\pm0.0155}_{\pm0.0244}$ & $0.2696^{\pm0.0146}_{\pm0.0216}$ & $0.2538^{\pm0.0142}_{\pm0.0273}$ & $0.2266^{\pm0.0136}_{\pm0.0154}$ \\
      $2.1500$ & $2.2170$ & $0.3629^{\pm0.0171}_{\pm0.0175}$ & $0.3142^{\pm0.0150}_{\pm0.0153}$ & $0.3271^{\pm0.0151}_{\pm0.0123}$ & $0.2915^{\pm0.0153}_{\pm0.0134}$ & $0.2232^{\pm0.0133}_{\pm0.0169}$ \\
      $2.2500$ & $2.2589$ & $0.3980^{\pm0.0236}_{\pm0.0401}$ & $0.3324^{\pm0.0200}_{\pm0.0431}$ & $0.2801^{\pm0.0175}_{\pm0.0170}$ & $0.2108^{\pm0.0130}_{\pm0.0119}$ & $0.1821^{\pm0.0115}_{\pm0.0107}$ \\
      $2.3500$ & $2.3001$ & $0.2993^{\pm0.0213}_{\pm0.0340}$ & $0.2929^{\pm0.0241}_{\pm0.0295}$ & $0.2948^{\pm0.0189}_{\pm0.0174}$ & $0.2476^{\pm0.0167}_{\pm0.0258}$ & $0.2237^{\pm0.0157}_{\pm0.0390}$ \\
      $2.4500$ & $2.3405$ & $0.4452^{\pm0.0287}_{\pm0.0388}$ & $0.4214^{\pm0.0275}_{\pm0.0676}$ & $0.4306^{\pm0.0275}_{\pm0.0245}$ & $0.3818^{\pm0.0240}_{\pm0.0336}$ & $0.3026^{\pm0.0210}_{\pm0.0303}$ \\
      $2.5500$ & $2.3803$ & $0.4814^{\pm0.0346}_{\pm0.1279}$ & $0.3609^{\pm0.0277}_{\pm0.0228}$ & $0.3652^{\pm0.0257}_{\pm0.0290}$ & $0.2821^{\pm0.0214}_{\pm0.0425}$ & $0.2793^{\pm0.0206}_{\pm0.0496}$ \\
      $2.6500$ & $2.4193$ & $0.5295^{\pm0.0494}_{\pm0.0261}$ & $0.3505^{\pm0.0338}_{\pm0.0150}$ & $0.3220^{\pm0.0306}_{\pm0.0140}$ & $0.3544^{\pm0.0339}_{\pm0.0160}$ & $0.2980^{\pm0.0284}_{\pm0.0154}$ \\
      $2.7500$ & $2.4578$ & $0.7359^{\pm0.0829}_{\pm0.0339}$ & $0.3617^{\pm0.0384}_{\pm0.0150}$ & $0.4093^{\pm0.0410}_{\pm0.0200}$ & $0.3380^{\pm0.0357}_{\pm0.0138}$ & $0.2924^{\pm0.0313}_{\pm0.0180}$ \\
      $2.8500$ & $2.4957$ & $0.4934^{\pm0.0515}_{\pm0.0245}$ & $0.3104^{\pm0.0379}_{\pm0.0148}$ & $0.3718^{\pm0.0448}_{\pm0.0209}$ & $0.3566^{\pm0.0392}_{\pm0.0198}$ & $0.2832^{\pm0.0304}_{\pm0.0148}$ \\
      $2.9500$ & $2.5330$ & $0.6718^{\pm0.1047}_{\pm0.0309}$ & $0.3422^{\pm0.0519}_{\pm0.0180}$ & $0.3600^{\pm0.0466}_{\pm0.0175}$ & $0.3196^{\pm0.0411}_{\pm0.0181}$ & $0.2948^{\pm0.0404}_{\pm0.0146}$ \\
      $3.0500$ & $2.5698$ & $0.7510^{\pm0.1210}_{\pm0.0503}$ & $0.3769^{\pm0.0651}_{\pm0.0246}$ & $0.4937^{\pm0.0637}_{\pm0.0419}$ & $0.4405^{\pm0.0638}_{\pm0.0294}$ & $0.3378^{\pm0.0488}_{\pm0.0336}$ \\
      $3.1500$ & $2.6061$ & $0.9479^{\pm0.1849}_{\pm0.0765}$ & $0.6846^{\pm0.1098}_{\pm0.0448}$ & $0.4364^{\pm0.0741}_{\pm0.0288}$ & $0.2288^{\pm0.0477}_{\pm0.0159}$ & $0.3663^{\pm0.0661}_{\pm0.0347}$ \\
      $3.2500$ & $2.6418$ & $0.8494^{\pm0.1894}_{\pm0.0923}$ & $0.6616^{\pm0.1241}_{\pm0.0579}$ & $0.4677^{\pm0.0921}_{\pm0.0461}$ & $0.3325^{\pm0.0787}_{\pm0.0329}$ & $0.3453^{\pm0.0741}_{\pm0.0323}$ \\
      $3.3500$ & $2.6771$ & $0.7554^{\pm0.3435}_{\pm0.0679}$ & $0.6573^{\pm0.1652}_{\pm0.0574}$ & $0.2406^{\pm0.1249}_{\pm0.0241}$ & $0.2626^{\pm0.0560}_{\pm0.0289}$ & $0.2138^{\pm0.0564}_{\pm0.0268}$ \\
      $3.4500$ & $2.7119$ & $0.7195^{\pm0.2952}_{\pm0.0885}$ & $0.9465^{\pm0.2898}_{\pm0.0981}$ & $0.5407^{\pm0.1511}_{\pm0.0586}$ & $0.3433^{\pm0.0745}_{\pm0.0332}$ & $0.2168^{\pm0.0950}_{\pm0.0227}$ \\
      $3.5500$ & $2.7463$ & $0.4325^{\pm0.1299}_{\pm0.0884}$ & $0.7958^{\pm0.1843}_{\pm0.1106}$ & $0.4842^{\pm0.1374}_{\pm0.0596}$ & $0.2155^{\pm0.1013}_{\pm0.0358}$ & $0.2609^{\pm0.0792}_{\pm0.0314}$ \\
\hline\hline
\end{tabular}
\label{table:table2}
\end{table*}
\end{center}

\begin{center}
\begin{table*}[h!]
	\caption{$d\sigma/dt$ $(\mu b/GeV^2)$ vs. $|t-t_{min}|$ $(GeV^2)$ for different photon beam energies. Superscripts are statistical errors and subscripts are systematic errors.}
	\begin{tabular}{|c|c|c|c|c|c|c|}
	\hline\hline
	$E_{\gamma}$ $(GeV)$ & $\sqrt{s}$ $(GeV)$ & \multicolumn{5}{|c|}{$|t-t_{min}|$ $(GeV^2)$} \\
	\cline{3-7}
	& & $0.460$ & $0.500$ & $0.540$ & $0.580$ & $0.620$ \\
        \hline
      $1.6500$ & $1.9942$ & $0.1025^{\pm0.0093}_{\pm0.0093}$ & $0.1043^{\pm0.0105}_{\pm0.0142}$ & $0.1058^{\pm0.0129}_{\pm0.0137}$ & $0.0624^{\pm0.0094}_{\pm0.0067}$ & $0.0521^{\pm0.0108}_{\pm0.0157}$ \\
      $1.7500$ & $2.0407$ & $0.1921^{\pm0.0101}_{\pm0.0188}$ & $0.1556^{\pm0.0092}_{\pm0.0093}$ & $0.1519^{\pm0.0096}_{\pm0.0137}$ & $0.1490^{\pm0.0097}_{\pm0.0200}$ & $0.1535^{\pm0.0156}_{\pm0.0131}$ \\
      $1.8500$ & $2.0861$ & $0.1575^{\pm0.0084}_{\pm0.0179}$ & $0.1516^{\pm0.0082}_{\pm0.0087}$ & $0.1552^{\pm0.0090}_{\pm0.0122}$ & $0.1350^{\pm0.0082}_{\pm0.0224}$ & $0.1201^{\pm0.0074}_{\pm0.0142}$ \\
      $1.9500$ & $2.1306$ & $0.1765^{\pm0.0100}_{\pm0.0214}$ & $0.1453^{\pm0.0085}_{\pm0.0074}$ & $0.1341^{\pm0.0086}_{\pm0.0315}$ & $0.1088^{\pm0.0073}_{\pm0.0134}$ & $0.1081^{\pm0.0071}_{\pm0.0115}$ \\
      $2.0500$ & $2.1742$ & $0.1849^{\pm0.0114}_{\pm0.0391}$ & $0.1535^{\pm0.0109}_{\pm0.0088}$ & $0.1453^{\pm0.0107}_{\pm0.0308}$ & $0.1302^{\pm0.0107}_{\pm0.0110}$ & $0.1100^{\pm0.0094}_{\pm0.0118}$ \\
      $2.1500$ & $2.2170$ & $0.0754^{\pm0.0214}_{\pm0.0088}$ & $0.0771^{\pm0.0272}_{\pm0.0090}$ & $0.0801^{\pm0.0285}_{\pm0.0139}$ & $0.0647^{\pm0.0293}_{\pm0.0112}$ & $0.0557^{\pm0.0249}_{\pm0.0180}$ \\
      $2.2500$ & $2.2589$ & $0.1956^{\pm0.0138}_{\pm0.0115}$ & $0.1759^{\pm0.0117}_{\pm0.0125}$ & $0.1774^{\pm0.0145}_{\pm0.0131}$ & $0.1411^{\pm0.0115}_{\pm0.0084}$ & $0.1334^{\pm0.0109}_{\pm0.0100}$ \\
      $2.3500$ & $2.3001$ & $0.1874^{\pm0.0135}_{\pm0.0324}$ & $0.1606^{\pm0.0126}_{\pm0.0149}$ & $0.1580^{\pm0.0118}_{\pm0.0298}$ & $0.1476^{\pm0.0146}_{\pm0.0369}$ & $0.1252^{\pm0.0101}_{\pm0.0086}$ \\
      $2.4500$ & $2.3405$ & $0.2775^{\pm0.0202}_{\pm0.0193}$ & $0.2731^{\pm0.0195}_{\pm0.0189}$ & $0.2614^{\pm0.0190}_{\pm0.0186}$ & $0.2566^{\pm0.0185}_{\pm0.0208}$ & $0.2175^{\pm0.0158}_{\pm0.0210}$ \\
      $2.6500$ & $2.4193$ & $0.3050^{\pm0.0296}_{\pm0.0156}$ & $0.3103^{\pm0.0309}_{\pm0.0156}$ & $0.2301^{\pm0.0239}_{\pm0.0114}$ & $0.1636^{\pm0.0172}_{\pm0.0090}$ & $0.1578^{\pm0.0171}_{\pm0.0115}$ \\
      $2.7500$ & $2.4578$ & $0.3154^{\pm0.0326}_{\pm0.0162}$ & $0.2764^{\pm0.0300}_{\pm0.0142}$ & $0.2483^{\pm0.0282}_{\pm0.0127}$ & $0.2049^{\pm0.0251}_{\pm0.0124}$ & $0.1663^{\pm0.0200}_{\pm0.0119}$ \\
      $2.8500$ & $2.4957$ & $0.2298^{\pm0.0276}_{\pm0.0118}$ & $0.1860^{\pm0.0244}_{\pm0.0096}$ & $0.2021^{\pm0.0263}_{\pm0.0103}$ & $0.1920^{\pm0.0245}_{\pm0.0103}$ & $0.1431^{\pm0.0201}_{\pm0.0108}$ \\
      $2.9500$ & $2.5330$ & $0.2963^{\pm0.0381}_{\pm0.0216}$ & $0.2257^{\pm0.0298}_{\pm0.0124}$ & $0.2635^{\pm0.0359}_{\pm0.0137}$ & $0.2299^{\pm0.0317}_{\pm0.0112}$ & $0.1699^{\pm0.0256}_{\pm0.0131}$ \\
      $3.0500$ & $2.5698$ & $0.2671^{\pm0.0454}_{\pm0.0261}$ & $0.2547^{\pm0.0375}_{\pm0.0244}$ & $0.1736^{\pm0.0295}_{\pm0.0189}$ & $0.1783^{\pm0.0321}_{\pm0.0174}$ & $0.1900^{\pm0.0308}_{\pm0.0172}$ \\
      $3.1500$ & $2.6061$ & $0.4051^{\pm0.0811}_{\pm0.0441}$ & $0.2319^{\pm0.0463}_{\pm0.0238}$ & $0.2694^{\pm0.0499}_{\pm0.0268}$ & $0.1827^{\pm0.0392}_{\pm0.0189}$ & $0.1828^{\pm0.0392}_{\pm0.0178}$ \\
      $3.2500$ & $2.6418$ & $0.3297^{\pm0.0666}_{\pm0.0269}$ & $0.2731^{\pm0.0508}_{\pm0.0218}$ & $0.3078^{\pm0.0617}_{\pm0.0289}$ & $0.2466^{\pm0.0506}_{\pm0.0203}$ & $0.1969^{\pm0.0479}_{\pm0.0174}$ \\
      $3.3500$ & $2.6771$ & $0.2348^{\pm0.0573}_{\pm0.0210}$ & $0.3011^{\pm0.0788}_{\pm0.0257}$ & $0.2931^{\pm0.0590}_{\pm0.0301}$ & $0.1544^{\pm0.0370}_{\pm0.0120}$ & $0.2338^{\pm0.0544}_{\pm0.0281}$ \\
      $3.4500$ & $2.7119$ & $0.2516^{\pm0.0827}_{\pm0.0260}$ & $0.3290^{\pm0.2141}_{\pm0.0511}$ & $0.2416^{\pm0.0732}_{\pm0.0270}$ & $0.2867^{\pm0.0804}_{\pm0.0304}$ & $0.1950^{\pm0.0496}_{\pm0.0261}$ \\
      $3.5500$ & $2.7463$ & $0.2255^{\pm0.1258}_{\pm0.0458}$ & $0.1608^{\pm0.0558}_{\pm0.0186}$ & $0.2188^{\pm0.0741}_{\pm0.0222}$ & $0.3006^{\pm0.0880}_{\pm0.0358}$ & $0.1310^{\pm0.0516}_{\pm0.0186}$ \\
\hline\hline
\end{tabular}
\label{table:table3}
\end{table*}
\end{center}

\begin{center}
\begin{table*}[h!]
	\caption{$d\sigma/dt$ $(\mu b/GeV^2)$ vs. $|t-t_{min}|$ $(GeV^2)$ for different photon beam energies. Superscripts are statistical errors and subscripts are systematic errors.}
	\begin{tabular}{|c|c|c|c|c|c|c|}
	\hline\hline
	$E_{\gamma}$ $(GeV)$ & $\sqrt{s}$ $(GeV)$ & \multicolumn{5}{|c|}{$|t-t_{min}|$ $(GeV^2)$} \\
	\cline{3-7}
	 & & $0.660$ & $0.700$ & $0.740$ & $0.780$ & $0.820$ \\
        \hline
     $1.6500$ & $1.9942$ & $0.0347^{\pm0.0119}_{\pm0.0067}$ &  -  &  -  & -  & -  \\
     $1.7500$ & $2.0407$ & $0.1480^{\pm0.0107}_{\pm0.0285}$ & $0.1423^{\pm0.0114}_{\pm0.0213}$ & $0.1288^{\pm0.0109}_{\pm0.0644}$ & $0.1195^{\pm0.0121}_{\pm0.0438}$ & $0.1290^{\pm0.0189}_{\pm0.0465}$ \\
     $1.8500$ & $2.0861$ & $0.1251^{\pm0.0083}_{\pm0.0124}$ & $0.1492^{\pm0.0100}_{\pm0.0177}$ & $0.1371^{\pm0.0085}_{\pm0.0092}$ & $0.1552^{\pm0.0117}_{\pm0.0500}$ & $0.1782^{\pm0.0142}_{\pm0.0232}$ \\
      $1.9500$ & $2.1306$ & $0.1114^{\pm0.0078}_{\pm0.0209}$ & $0.1108^{\pm0.0091}_{\pm0.0114}$ & $0.0958^{\pm0.0072}_{\pm0.0263}$ & $0.0988^{\pm0.0080}_{\pm0.0176}$ & $0.1141^{\pm0.0094}_{\pm0.0083}$ \\
      $2.0500$ & $2.1742$ & $0.0844^{\pm0.0075}_{\pm0.0064}$ & $0.0757^{\pm0.0075}_{\pm0.0130}$ & $0.0807^{\pm0.0071}_{\pm0.0078}$ & $0.0687^{\pm0.0067}_{\pm0.0226}$ & $0.0780^{\pm0.0083}_{\pm0.0126}$ \\
      $2.1500$ & $2.2170$ & $0.1362^{\pm0.0114}_{\pm0.0097}$ & $0.1118^{\pm0.0100}_{\pm0.0111}$ & $0.0924^{\pm0.0076}_{\pm0.0289}$ & $0.1010^{\pm0.0102}_{\pm0.0128}$ & $0.0853^{\pm0.0081}_{\pm0.0074}$ \\
      $2.2500$ & $2.2589$ & $0.1288^{\pm0.0118}_{\pm0.0091}$ & $0.1255^{\pm0.0118}_{\pm0.0084}$ & $0.1102^{\pm0.0091}_{\pm0.0091}$ & $0.0913^{\pm0.0100}_{\pm0.0133}$ & $0.0864^{\pm0.0088}_{\pm0.0253}$ \\
      $2.3500$ & $2.3001$ & $0.1301^{\pm0.0146}_{\pm0.0099}$ & $0.1307^{\pm0.0116}_{\pm0.0214}$ & $0.1129^{\pm0.0133}_{\pm0.0264}$ & $0.0826^{\pm0.0083}_{\pm0.0081}$ & $0.0791^{\pm0.0100}_{\pm0.0096}$ \\
      $2.4500$ & $2.3405$ & $0.2154^{\pm0.0153}_{\pm0.0191}$ & $0.1804^{\pm0.0139}_{\pm0.0144}$ & $0.1833^{\pm0.0142}_{\pm0.0170}$ & $0.1487^{\pm0.0126}_{\pm0.0146}$ & $0.1444^{\pm0.0127}_{\pm0.0221}$ \\
      $2.5500$ & $2.3803$ & $0.1337^{\pm0.0112}_{\pm0.0161}$ & $0.1137^{\pm0.0097}_{\pm0.0134}$ & $0.1098^{\pm0.0096}_{\pm0.0130}$ & $0.0964^{\pm0.0090}_{\pm0.0227}$ & $0.0921^{\pm0.0089}_{\pm0.0123}$ \\
      $2.6500$ & $2.4193$ & $0.1501^{\pm0.0179}_{\pm0.0104}$ & $0.1064^{\pm0.0140}_{\pm0.0088}$ & $0.1117^{\pm0.0132}_{\pm0.0081}$ & $0.1043^{\pm0.0134}_{\pm0.0089}$ & $0.1275^{\pm0.0160}_{\pm0.0106}$ \\
      $2.7500$ & $2.4578$ & $0.1464^{\pm0.0183}_{\pm0.0103}$ & $0.1325^{\pm0.0176}_{\pm0.0097}$ & $0.1176^{\pm0.0163}_{\pm0.0095}$ & $0.1100^{\pm0.0170}_{\pm0.0077}$ & $0.0804^{\pm0.0136}_{\pm0.0083}$ \\
      $2.8500$ & $2.4957$ & $0.1594^{\pm0.0215}_{\pm0.0113}$ & $0.1415^{\pm0.0216}_{\pm0.0106}$ & $0.1195^{\pm0.0194}_{\pm0.0086}$ & $0.0980^{\pm0.0162}_{\pm0.0074}$ & $0.0795^{\pm0.0151}_{\pm0.0086}$ \\
      $2.9500$ & $2.5330$ & $0.1597^{\pm0.0253}_{\pm0.0113}$ & $0.1309^{\pm0.0228}_{\pm0.0099}$ & $0.1271^{\pm0.0228}_{\pm0.0095}$ & $0.1057^{\pm0.0201}_{\pm0.0078}$ & $0.0932^{\pm0.0184}_{\pm0.0098}$ \\
      $3.0500$ & $2.5698$ & $0.1544^{\pm0.0268}_{\pm0.0168}$ & $0.1012^{\pm0.0209}_{\pm0.0092}$ & $0.1441^{\pm0.0259}_{\pm0.0125}$ & $0.1278^{\pm0.0284}_{\pm0.0113}$ & $0.1289^{\pm0.0258}_{\pm0.0216}$ \\
      $3.1500$ & $2.6061$ & $0.1687^{\pm0.0394}_{\pm0.0156}$ & $0.1299^{\pm0.0316}_{\pm0.0121}$ & $0.1519^{\pm0.0330}_{\pm0.0133}$ & $0.1769^{\pm0.0374}_{\pm0.0157}$ & $0.1095^{\pm0.0278}_{\pm0.0178}$ \\
      $3.2500$ & $2.6418$ & $0.1808^{\pm0.0454}_{\pm0.0171}$ & $0.2395^{\pm0.0493}_{\pm0.0224}$ & $0.1369^{\pm0.0310}_{\pm0.0128}$ & $0.1210^{\pm0.0293}_{\pm0.0106}$ & $0.1115^{\pm0.0280}_{\pm0.0164}$ \\
      $3.3500$ & $2.6771$ & $0.2080^{\pm0.0544}_{\pm0.0202}$ & $0.1995^{\pm0.0517}_{\pm0.0194}$ & $0.1535^{\pm0.0430}_{\pm0.0134}$ & $0.1289^{\pm0.0320}_{\pm0.0145}$ & $0.1217^{\pm0.0328}_{\pm0.0181}$ \\
      $3.4500$ & $2.7119$ & $0.1865^{\pm0.0597}_{\pm0.0264}$ & $0.2104^{\pm0.0600}_{\pm0.0281}$ & $0.1438^{\pm0.0474}_{\pm0.0193}$ & $0.1405^{\pm0.0415}_{\pm0.0206}$ & $0.1433^{\pm0.0436}_{\pm0.0323}$ \\
      $3.5500$ & $2.7463$ & $0.1309^{\pm0.0468}_{\pm0.0194}$ & $0.2009^{\pm0.0670}_{\pm0.0489}$ & $0.1237^{\pm0.0397}_{\pm0.0164}$ & $0.0870^{\pm0.0350}_{\pm0.0119}$ & $0.1018^{\pm0.0368}_{\pm0.0245}$ \\
	\hline\hline
\end{tabular}
\label{table:table4}
\end{table*}
\end{center}

\begin{center}
\begin{table*}[h!]
	\caption{$d\sigma/dt$ $(\mu b/GeV^2)$ vs. $|t-t_{min}|$ $(GeV^2)$ for different photon beam energies. Superscripts are statistical errors and subscripts are systematic errors.}
	\begin{tabular}{|c|c|c|c|c|c|c|}
	\hline\hline
	$E_{\gamma}$ $(GeV)$ & $\sqrt{s}$ $(GeV)$ & \multicolumn{5}{|c|}{$|t-t_{min}|$ $(GeV^2)$} \\
	\cline{3-7}
	 & & $0.860$ & $0.900$ & $0.940$ & $0.980$ & $1.020$ \\
        \hline
      $1.7500$ & $2.0407$ & $0.0649^{\pm0.0067}_{\pm0.0152}$ & $0.0187^{\pm0.0027}_{\pm0.0053}$ & - & - & - \\
      $1.8500$ & $2.0861$ & $0.1699^{\pm0.0258}_{\pm0.0531}$ & $0.1681^{\pm0.0285}_{\pm0.0307}$ & $0.1855^{\pm0.0391}_{\pm0.0118}$ & $0.1666^{\pm0.0585}_{\pm0.0117}$ & $0.1670^{\pm0.0617}_{\pm0.0118}$ \\
      $1.9500$ & $2.1306$ & $0.1294^{\pm0.0124}_{\pm0.0099}$ & $0.1405^{\pm0.0133}_{\pm0.0224}$ & $0.1427^{\pm0.0137}_{\pm0.0091}$ & $0.1591^{\pm0.0168}_{\pm0.0112}$ & $0.1567^{\pm0.0177}_{\pm0.0123}$ \\
      $2.0500$ & $2.1742$ & $0.0691^{\pm0.0065}_{\pm0.0195}$ & $0.0738^{\pm0.0071}_{\pm0.0107}$ & $0.0800^{\pm0.0078}_{\pm0.0208}$ & $0.0837^{\pm0.0083}_{\pm0.0095}$ & $0.0769^{\pm0.0080}_{\pm0.0087}$ \\
      $2.1500$ & $2.2170$ & $0.0795^{\pm0.0078}_{\pm0.0068}$ & $0.0808^{\pm0.0086}_{\pm0.0126}$ & $0.0705^{\pm0.0075}_{\pm0.0092}$ & $0.0641^{\pm0.0073}_{\pm0.0072}$ & $0.0620^{\pm0.0074}_{\pm0.0071}$ \\
      $2.2500$ & $2.2589$ & $0.0775^{\pm0.0081}_{\pm0.0174}$ & $0.0740^{\pm0.0083}_{\pm0.0203}$ & $0.0726^{\pm0.0089}_{\pm0.0069}$ & $0.0718^{\pm0.0089}_{\pm0.0210}$ & $0.0664^{\pm0.0099}_{\pm0.0095}$ \\
      $2.3500$ & $2.3001$ & $0.0786^{\pm0.0092}_{\pm0.0280}$ & $0.0841^{\pm0.0096}_{\pm0.0130}$ & $0.0698^{\pm0.0093}_{\pm0.0325}$ & $0.0690^{\pm0.0096}_{\pm0.0153}$ & $0.0680^{\pm0.0096}_{\pm0.0097}$ \\
      $2.4500$ & $2.3405$ & $0.1335^{\pm0.0118}_{\pm0.0127}$ & $0.1178^{\pm0.0109}_{\pm0.0289}$ & $0.1134^{\pm0.0112}_{\pm0.0351}$ & $0.0888^{\pm0.0109}_{\pm0.0311}$ & $0.0678^{\pm0.0095}_{\pm0.0108}$ \\
      $2.5500$ & $2.3803$ & $0.0799^{\pm0.0082}_{\pm0.0131}$ & $0.0790^{\pm0.0083}_{\pm0.0095}$ & $0.0747^{\pm0.0083}_{\pm0.0145}$ & $0.0698^{\pm0.0092}_{\pm0.0174}$ & $0.0658^{\pm0.0094}_{\pm0.0105}$ \\
      $2.6500$ & $2.4193$ & $0.0940^{\pm0.0129}_{\pm0.0076}$ & $0.0701^{\pm0.0112}_{\pm0.0053}$ & $0.0798^{\pm0.0125}_{\pm0.0060}$ & $0.0692^{\pm0.0115}_{\pm0.0053}$ & $0.0563^{\pm0.0103}_{\pm0.0067}$ \\
      $2.7500$ & $2.4578$ & $0.1119^{\pm0.0183}_{\pm0.0086}$ & $0.0832^{\pm0.0145}_{\pm0.0067}$ & $0.0938^{\pm0.0165}_{\pm0.0071}$ & $0.0690^{\pm0.0133}_{\pm0.0064}$ & $0.0770^{\pm0.0147}_{\pm0.0091}$ \\
      $2.8500$ & $2.4957$ & $0.0815^{\pm0.0158}_{\pm0.0086}$ & $0.0661^{\pm0.0140}_{\pm0.0073}$ & $0.0644^{\pm0.0138}_{\pm0.0075}$ & $0.0733^{\pm0.0158}_{\pm0.0087}$ & $0.0464^{\pm0.0112}_{\pm0.0058}$ \\
      $2.9500$ & $2.5330$ & $0.0777^{\pm0.0168}_{\pm0.0083}$ & $0.0864^{\pm0.0194}_{\pm0.0092}$ & $0.0635^{\pm0.0160}_{\pm0.0069}$ & $0.0618^{\pm0.0145}_{\pm0.0072}$ & $0.0458^{\pm0.0129}_{\pm0.0059}$ \\
      $3.0500$ & $2.5698$ & $0.0527^{\pm0.0138}_{\pm0.0086}$ & $0.0705^{\pm0.0187}_{\pm0.0117}$ & $0.0398^{\pm0.0121}_{\pm0.0066}$ & $0.0481^{\pm0.0136}_{\pm0.0082}$ & $0.0687^{\pm0.0197}_{\pm0.0134}$ \\
      $3.1500$ & $2.6061$ & $0.1142^{\pm0.0318}_{\pm0.0187}$ & $0.0564^{\pm0.0188}_{\pm0.0105}$ & $0.0836^{\pm0.0233}_{\pm0.0137}$ & $0.0509^{\pm0.0174}_{\pm0.0083}$ & $0.0499^{\pm0.0172}_{\pm0.0092}$ \\
      $3.2500$ & $2.6418$ & $0.0760^{\pm0.0202}_{\pm0.0112}$ & $0.0747^{\pm0.0241}_{\pm0.0118}$ & $0.0848^{\pm0.0236}_{\pm0.0125}$ & $0.0650^{\pm0.0205}_{\pm0.0096}$ & $0.0437^{\pm0.0169}_{\pm0.0100}$ \\
      $3.3500$ & $2.6771$ & $0.0799^{\pm0.0248}_{\pm0.0118}$ & $0.0807^{\pm0.0260}_{\pm0.0143}$ & $0.0924^{\pm0.0287}_{\pm0.0136}$ & $0.0819^{\pm0.0281}_{\pm0.0125}$ & $0.0598^{\pm0.0237}_{\pm0.0137}$ \\
      $3.4500$ & $2.7119$ & $0.0831^{\pm0.0301}_{\pm0.0184}$ & $0.1055^{\pm0.0371}_{\pm0.0235}$ & $0.0951^{\pm0.0393}_{\pm0.0211}$ & $0.0852^{\pm0.0322}_{\pm0.0194}$ & $0.0674^{\pm0.0298}_{\pm0.0287}$ \\
      $3.5500$ & $2.7463$ & $0.1057^{\pm0.0391}_{\pm0.0233}$ & $0.0847^{\pm0.0357}_{\pm0.0187}$ & $0.0577^{\pm0.0294}_{\pm0.0127}$ & $0.0453^{\pm0.0224}_{\pm0.0100}$ & $0.0575^{\pm0.0263}_{\pm0.0258}$ \\
	\hline\hline
\end{tabular}
\label{table:table5}
\end{table*}
\end{center}

\begin{center}
\begin{table*}[h!]
	\caption{$d\sigma/dt$ $(\mu b/GeV^2)$ vs. $|t-t_{min}|$ $(GeV^2)$ for different photon beam energies. Superscripts are statistical errors and subscripts are systematic errors.}
	\begin{tabular}{|c|c|c|c|c|c|c|}
	\hline\hline
	$E_{\gamma}$ $(GeV)$ & $\sqrt{s}$ $(GeV)$ & \multicolumn{5}{|c|}{$|t-t_{min}|$ $(GeV^2)$} \\
	\cline{3-7}
	 &  & $1.060$ & $1.100$ & $1.140$ & $1.180$ & $1.220$ \\
        \hline
      $1.8500$ & $2.0861$ & $0.1411^{\pm0.0481}_{\pm0.0113}$ & $0.0672^{\pm0.0264}_{\pm0.0054}$ & - & - & - \\
      $1.9500$ & $2.1306$ & $0.1671^{\pm0.0260}_{\pm0.0134}$ & $0.1558^{\pm0.0252}_{\pm0.0138}$ & $0.1311^{\pm0.0295}_{\pm0.0148}$ & $0.1175^{\pm0.0407}_{\pm0.0135}$ & $0.1222^{\pm0.0400}_{\pm0.0231}$ \\
      $2.0500$ & $2.1742$ & $0.0818^{\pm0.0092}_{\pm0.0110}$ & $0.0859^{\pm0.0099}_{\pm0.0118}$ & $0.0902^{\pm0.0108}_{\pm0.0111}$ & $0.0933^{\pm0.0115}_{\pm0.0116}$ & $0.1138^{\pm0.0153}_{\pm0.0156}$ \\
      $2.1500$ & $2.2170$ & $0.0655^{\pm0.0080}_{\pm0.0088}$ & $0.0605^{\pm0.0076}_{\pm0.0082}$ & $0.0606^{\pm0.0077}_{\pm0.0075}$ & $0.0589^{\pm0.0078}_{\pm0.0073}$ & $0.0546^{\pm0.0077}_{\pm0.0075}$ \\
      $2.2500$ & $2.2589$ & $0.0707^{\pm0.0097}_{\pm0.0144}$ & $0.0627^{\pm0.0089}_{\pm0.0130}$ & $0.0564^{\pm0.0082}_{\pm0.0113}$ & $0.0541^{\pm0.0076}_{\pm0.0108}$ & $0.0546^{\pm0.0079}_{\pm0.0099}$ \\
      $2.3500$ & $2.3001$ & $0.0587^{\pm0.0088}_{\pm0.0120}$ & $0.0522^{\pm0.0087}_{\pm0.0107}$ & $0.0479^{\pm0.0070}_{\pm0.0096}$ & $0.0415^{\pm0.0070}_{\pm0.0083}$ & $0.0414^{\pm0.0073}_{\pm0.0075}$ \\
      $2.4500$ & $2.3405$ & $0.0611^{\pm0.0086}_{\pm0.0106}$ & $0.0624^{\pm0.0088}_{\pm0.0108}$ & $0.0598^{\pm0.0079}_{\pm0.0105}$ & $0.0639^{\pm0.0082}_{\pm0.0126}$ & $0.0567^{\pm0.0080}_{\pm0.0118}$ \\
      $2.5500$ & $2.3803$ & $0.0554^{\pm0.0083}_{\pm0.0101}$ & $0.0521^{\pm0.0084}_{\pm0.0096}$ & $0.0505^{\pm0.0079}_{\pm0.0099}$ & $0.0479^{\pm0.0075}_{\pm0.0093}$ & $0.0410^{\pm0.0067}_{\pm0.0085}$ \\
      $2.6500$ & $2.4193$ & $0.0675^{\pm0.0128}_{\pm0.0080}$ & $0.0568^{\pm0.0117}_{\pm0.0067}$ & $0.0515^{\pm0.0106}_{\pm0.0061}$ & $0.0597^{\pm0.0132}_{\pm0.0070}$ & $0.0499^{\pm0.0125}_{\pm0.0075}$ \\
      $2.7500$ & $2.4578$ & $0.0594^{\pm0.0123}_{\pm0.0073}$ & $0.0542^{\pm0.0124}_{\pm0.0064}$ & $0.0528^{\pm0.0121}_{\pm0.0063}$ & $0.0423^{\pm0.0120}_{\pm0.0051}$ & $0.0300^{\pm0.0086}_{\pm0.0045}$ \\
      $2.8500$ & $2.4957$ & $0.0483^{\pm0.0114}_{\pm0.0069}$ & $0.0398^{\pm0.0105}_{\pm0.0051}$ & $0.0464^{\pm0.0134}_{\pm0.0058}$ & $0.0442^{\pm0.0142}_{\pm0.0055}$ & $0.0587^{\pm0.0176}_{\pm0.0093}$ \\
      $2.9500$ & $2.5330$ & $0.0570^{\pm0.0158}_{\pm0.0072}$ & $0.0532^{\pm0.0150}_{\pm0.0068}$ & $0.0448^{\pm0.0155}_{\pm0.0056}$ & $0.0463^{\pm0.0145}_{\pm0.0059}$ & $0.0417^{\pm0.0165}_{\pm0.0065}$ \\
      $3.0500$ & $2.5698$ & $0.0809^{\pm0.0244}_{\pm0.0144}$ & $0.0495^{\pm0.0150}_{\pm0.0089}$ & $0.0642^{\pm0.0207}_{\pm0.0119}$ & $0.0500^{\pm0.0190}_{\pm0.0089}$ & $0.0741^{\pm0.0280}_{\pm0.0239}$ \\
      $3.1500$ & $2.6061$ & $0.0585^{\pm0.0209}_{\pm0.0106}$ & $0.0608^{\pm0.0217}_{\pm0.0116}$ & $0.0644^{\pm0.0243}_{\pm0.0116}$ & $0.0603^{\pm0.0254}_{\pm0.0109}$ & $0.0600^{\pm0.0277}_{\pm0.0202}$ \\
      $3.2500$ & $2.6418$ & $0.0623^{\pm0.0232}_{\pm0.0143}$ & $0.0560^{\pm0.0218}_{\pm0.0130}$ & $0.0430^{\pm0.0172}_{\pm0.0099}$ & $0.0518^{\pm0.0235}_{\pm0.0120}$ & $0.0506^{\pm0.0237}_{\pm0.0274}$ \\
      $3.3500$ & $2.6771$ & $0.0457^{\pm0.0208}_{\pm0.0105}$ & $0.0635^{\pm0.0244}_{\pm0.0148}$ & $0.0462^{\pm0.0224}_{\pm0.0106}$ & $0.0285^{\pm0.0138}_{\pm0.0066}$ & $0.0609^{\pm0.0333}_{\pm0.0332}$ \\
      $3.4500$ & $2.7119$ & $0.0856^{\pm0.0325}_{\pm0.0366}$ & $0.0538^{\pm0.0215}_{\pm0.0228}$ & $0.0362^{\pm0.0179}_{\pm0.0156}$ & $0.0522^{\pm0.0263}_{\pm0.0223}$ & $0.0325^{\pm0.0190}_{\pm0.0220}$ \\
      $3.5500$ & $2.7463$ & $0.0439^{\pm0.0248}_{\pm0.0188}$ & $0.0473^{\pm0.0246}_{\pm0.0202}$ & $0.0266^{\pm0.0174}_{\pm0.0113}$ & $0.0564^{\pm0.0252}_{\pm0.0239}$ & $0.0453^{\pm0.0326}_{\pm0.0307}$ \\
	\hline
\hline
\end{tabular}
\label{table:table6}
\end{table*}
\end{center}

\begin{center}
\begin{table*}[h!]
	\caption{$d\sigma/dt$ $(\mu b/GeV^2)$ vs. $|t-t_{min}|$ $(GeV^2)$ for different photon beam energies. Superscripts are statistical errors and subscripts are systematic errors.}
	\begin{tabular}{|c|c|c|c|c|c|c|}
	\hline\hline
	$E_{\gamma}$ $(GeV)$ & $\sqrt{s}$ $(GeV)$ & \multicolumn{5}{|c|}{$|t-t_{min}|$ $(GeV^2)$} \\
	\cline{3-7}
	& & $1.260$ & $1.300$ & $1.340$ & $1.380$ & $1.420$ \\
        \hline
      $1.9500$ & $2.1306$ & $0.1042^{\pm0.0378}_{\pm0.0199}$ & $0.0473^{\pm0.0181}_{\pm0.0109}$ & $0.0241^{\pm0.0049}_{\pm0.0056}$ & $0.0279^{\pm0.0061}_{\pm0.0206}$ &$0.0074^{\pm0.0026}_{\pm0.0054}$ \\
      $2.0500$ & $2.1742$ & $0.0959^{\pm0.0191}_{\pm0.0131}$ & $0.1046^{\pm0.0242}_{\pm0.0140}$ & $0.0890^{\pm0.0280}_{\pm0.0120}$ & $0.1034^{\pm0.0351}_{\pm0.0150}$ & $0.0860^{\pm0.0310}_{\pm0.0125}$ \\
      $2.1500$ & $2.2170$ & $0.0642^{\pm0.0092}_{\pm0.0088}$ & $0.0669^{\pm0.0104}_{\pm0.0090}$ & $0.0680^{\pm0.0105}_{\pm0.0091}$ & $0.0787^{\pm0.0133}_{\pm0.0114}$ &n$0.0855^{\pm0.0182}_{\pm0.0125}$ \\
      $2.2500$ & $2.2589$ & $0.0519^{\pm0.0078}_{\pm0.0094}$ & $0.0596^{\pm0.0085}_{\pm0.0086}$ & $0.0497^{\pm0.0077}_{\pm0.0073}$ & $0.0526^{\pm0.0084}_{\pm0.0088}$ & $0.0663^{\pm0.0114}_{\pm0.0111}$ \\
      $2.3500$ & $2.3001$ & $0.0508^{\pm0.0087}_{\pm0.0093}$ & $0.0458^{\pm0.0081}_{\pm0.0066}$ & $0.0413^{\pm0.0075}_{\pm0.0062}$ & $0.0407^{\pm0.0076}_{\pm0.0068}$ & $0.0399^{\pm0.0077}_{\pm0.0067}$ \\
      $2.4500$ & $2.3405$ & $0.0535^{\pm0.0074}_{\pm0.0112}$ & $0.0511^{\pm0.0078}_{\pm0.0134}$ & $0.0443^{\pm0.0069}_{\pm0.0117}$ & $0.0464^{\pm0.0071}_{\pm0.0204}$ & $0.0431^{\pm0.0071}_{\pm0.0190}$ \\
      $2.5500$ & $2.3803$ & $0.0419^{\pm0.0072}_{\pm0.0088}$ & $0.0397^{\pm0.0068}_{\pm0.0104}$ & $0.0257^{\pm0.0049}_{\pm0.0067}$ & $0.0234^{\pm0.0044}_{\pm0.0103}$ & $0.0190^{\pm0.0040}_{\pm0.0083}$ \\
      $2.6500$ & $2.4193$ & $0.0355^{\pm0.0093}_{\pm0.0053}$ & $0.0357^{\pm0.0099}_{\pm0.0056}$ & $0.0308^{\pm0.0099}_{\pm0.0046}$ & $0.0318^{\pm0.0104}_{\pm0.0048}$ & $0.0225^{\pm0.0081}_{\pm0.0083}$ \\
      $2.7500$ & $2.4578$ & $0.0365^{\pm0.0114}_{\pm0.0055}$ & $0.0432^{\pm0.0144}_{\pm0.0065}$ & $0.0337^{\pm0.0126}_{\pm0.0050}$ & $0.0386^{\pm0.0142}_{\pm0.0061}$ & $0.0522^{\pm0.0196}_{\pm0.0194}$ \\
      $2.8500$ & $2.4957$ & $0.0527^{\pm0.0175}_{\pm0.0084}$ & $0.0576^{\pm0.0196}_{\pm0.0090}$ & $0.0351^{\pm0.0150}_{\pm0.0055}$ & $0.0452^{\pm0.0182}_{\pm0.0071}$ & $0.0530^{\pm0.0243}_{\pm0.0097}$ \\
      $2.9500$ & $2.5330$ & $0.0449^{\pm0.0163}_{\pm0.0075}$ & $0.0461^{\pm0.0195}_{\pm0.0072}$ & $0.0283^{\pm0.0110}_{\pm0.0044}$ & $0.0529^{\pm0.0265}_{\pm0.0084}$ & $0.0453^{\pm0.0225}_{\pm0.0083}$ \\
      $3.0500$ & $2.5698$ & $0.0614^{\pm0.0232}_{\pm0.0197}$ & $0.0704^{\pm0.0289}_{\pm0.0227}$ & $0.0626^{\pm0.0296}_{\pm0.0201}$ & $0.0417^{\pm0.0192}_{\pm0.0134}$ & $0.0362^{\pm0.0178}_{\pm0.0128}$ \\
      $3.1500$ & $2.6061$ & $0.0482^{\pm0.0225}_{\pm0.0155}$ & $0.0667^{\pm0.0336}_{\pm0.0214}$ & $0.0594^{\pm0.0280}_{\pm0.0192}$ & $0.0431^{\pm0.0231}_{\pm0.0139}$ & $0.0606^{\pm0.0292}_{\pm0.0215}$ \\
      $3.2500$ & $2.6418$ & $0.0324^{\pm0.0150}_{\pm0.0177}$ & $0.0345^{\pm0.0181}_{\pm0.0188}$ & $0.0388^{\pm0.0213}_{\pm0.0212}$ & $0.0305^{\pm0.0175}_{\pm0.0165}$ & $0.0314^{\pm0.0170}_{\pm0.0190}$ \\
      $3.3500$ & $2.6771$ & $0.0491^{\pm0.0255}_{\pm0.0272}$ & $0.0499^{\pm0.0295}_{\pm0.0272}$ & $0.0565^{\pm0.0301}_{\pm0.0306}$ & $0.0344^{\pm0.0212}_{\pm0.0187}$ & $0.0251^{\pm0.0180}_{\pm0.0153}$ \\
      $3.4500$ & $2.7119$ & $0.0194^{\pm0.0110}_{\pm0.0133}$ & $0.0264^{\pm0.0165}_{\pm0.0179}$ & $0.0389^{\pm0.0212}_{\pm0.0265}$ & $0.0111^{\pm0.0087}_{\pm0.0075}$ & $0.0304^{\pm0.0239}_{\pm0.0196}$ \\
      $3.5500$ & $2.7463$ & $0.0254^{\pm0.0177}_{\pm0.0172}$ & $0.0136^{\pm0.0123}_{\pm0.0092}$ & $0.0300^{\pm0.0247}_{\pm0.0203}$ & $0.0152^{\pm0.0167}_{\pm0.0103}$ & $0.0320^{\pm0.0299}_{\pm0.0224}$ \\
	\hline
\hline
\end{tabular}
\label{table:table7}
\end{table*}
\end{center}

\begin{center}
\begin{table*}[h!]
	\caption{$d\sigma/dt$ $(\mu b/GeV^2)$ vs. $|t-t_{min}|$ $(GeV^2)$ for different photon beam energies. Superscripts are statistical errors and subscripts are systematic errors.}
	\begin{tabular}{|c|c|c|c|c|c|c|}
	\hline\hline
	$E_{\gamma}$ $(GeV)$ & $\sqrt{s}$ $(GeV)$ & \multicolumn{5}{|c|}{$|t-t_{min}|$ $(GeV^2)$} \\
	\cline{3-7}
	& & $1.460$ & $1.500$ & $1.540$ & $1.580$ & $1.620$ \\
        \hline
      $2.0500$ & $2.1742$ & $0.0519^{\pm0.0204}_{\pm0.0060}$ & $0.0232^{\pm0.0110}_{\pm0.0027}$ & $0.0157^{\pm0.0085}_{\pm0.0027}$ & - & - \\
      $2.1500$ & $2.2170$ & $0.0754^{\pm0.0214}_{\pm0.0088}$ & $0.0771^{\pm0.0272}_{\pm0.0090}$ & $0.0801^{\pm0.0285}_{\pm0.0139}$ & $0.0647^{\pm0.0293}_{\pm0.0112}$ & $0.0557^{\pm0.0249}_{\pm0.0180}$ \\
      $2.2500$ & $2.2589$ & $0.0641^{\pm0.0111}_{\pm0.0129}$ & $0.0545^{\pm0.0101}_{\pm0.0110}$ & $0.0684^{\pm0.0143}_{\pm0.0149}$ & $0.0979^{\pm0.0211}_{\pm0.0206}$ & $0.0796^{\pm0.0227}_{\pm0.0174}$ \\
      $2.3500$ & $2.3001$ & $0.0349^{\pm0.0069}_{\pm0.0070}$ & $0.0360^{\pm0.0075}_{\pm0.0073}$ & $0.0359^{\pm0.0076}_{\pm0.0078}$ & $0.0375^{\pm0.0103}_{\pm0.0083}$ & $0.0312^{\pm0.0100}_{\pm0.0068}$ \\
      $2.4500$ & $2.3405$ & $0.0412^{\pm0.0072}_{\pm0.0156}$ & $0.0418^{\pm0.0067}_{\pm0.0158}$ & $0.0412^{\pm0.0068}_{\pm0.0131}$ & $0.0440^{\pm0.0066}_{\pm0.0139}$ & $0.0351^{\pm0.0061}_{\pm0.0144}$ \\
      $2.5500$ & $2.3803$ & $0.0269^{\pm0.0054}_{\pm0.0102}$ & $0.0308^{\pm0.0062}_{\pm0.0117}$ & $0.0329^{\pm0.0065}_{\pm0.0105}$ & $0.0461^{\pm0.0090}_{\pm0.0147}$ & $0.0381^{\pm0.0078}_{\pm0.0157}$ \\
      $2.6500$ & $2.4193$ & $0.0345^{\pm0.0130}_{\pm0.0128}$ & $0.0444^{\pm0.0177}_{\pm0.0165}$ & $0.0350^{\pm0.0155}_{\pm0.0130}$ & $0.0313^{\pm0.0149}_{\pm0.0116}$ & $0.0361^{\pm0.0171}_{\pm0.0116}$ \\
      $2.7500$ & $2.4578$ & $0.0311^{\pm0.0137}_{\pm0.0115}$ & $0.0326^{\pm0.0143}_{\pm0.0122}$ & $0.0335^{\pm0.0169}_{\pm0.0129}$ & $0.0269^{\pm0.0130}_{\pm0.0102}$ & $0.0242^{\pm0.0147}_{\pm0.0078}$ \\
      $2.8500$ & $2.4957$ & $0.0501^{\pm0.0235}_{\pm0.0092}$ & $0.0500^{\pm0.0257}_{\pm0.0091}$ & $0.0381^{\pm0.0220}_{\pm0.0070}$ & $0.0449^{\pm0.0296}_{\pm0.0082}$ & - \\
      $2.9500$ & $2.5330$ & $0.0402^{\pm0.0222}_{\pm0.0073}$ & $0.0413^{\pm0.0236}_{\pm0.0076}$ & $0.0309^{\pm0.0193}_{\pm0.0056}$ & $0.0464^{\pm0.0344}_{\pm0.0089}$ & - \\
      $3.0500$ & $2.5698$ & $0.0352^{\pm0.0183}_{\pm0.0124}$ & $0.0297^{\pm0.0205}_{\pm0.0105}$ & $0.0450^{\pm0.0319}_{\pm0.0159}$ & $0.0431^{\pm0.0354}_{\pm0.0152}$ & - \\
      $3.1500$ & $2.6061$ & $0.0706^{\pm0.0392}_{\pm0.0250}$ & $0.0269^{\pm0.0201}_{\pm0.0095}$ & $0.0505^{\pm0.0390}_{\pm0.0179}$ & $0.0321^{\pm0.0252}_{\pm0.0113}$ & - \\
      $3.2500$ & $2.6418$ & $0.0357^{\pm0.0226}_{\pm0.0217}$ & $0.0270^{\pm0.0198}_{\pm0.0164}$ & $0.0342^{\pm0.0235}_{\pm0.0207}$ & $0.0092^{\pm0.0082}_{\pm0.0055}$ & - \\
      $3.3500$ & $2.6771$ & $0.0307^{\pm0.0231}_{\pm0.0186}$ & $0.0319^{\pm0.0251}_{\pm0.0193}$ & $0.0287^{\pm0.0218}_{\pm0.0173}$ & $0.0241^{\pm0.0195}_{\pm0.0146}$ & - \\
      $3.4500$ & $2.7119$ & $0.0416^{\pm0.0339}_{\pm0.0291}$ & $0.0403^{\pm0.0348}_{\pm0.0323}$ & $0.0313^{\pm0.0311}_{\pm0.0219}$ & $0.0326^{\pm0.0293}_{\pm0.0260}$ & - \\
      $3.5500$ & $2.7463$ & $0.0106^{\pm0.0108}_{\pm0.0085}$ & $0.0178^{\pm0.0175}_{\pm0.0142}$ & $0.0175^{\pm0.0184}_{\pm0.0140}$ & $0.0325^{\pm0.0324}_{\pm0.0234}$ & - \\
	\hline
\hline
\end{tabular}
\label{table:table8}
\end{table*}
\end{center}

\begin{center}
\begin{table*}[h!]
	\caption{$d\sigma/dt$ $(\mu b/GeV^2)$ vs. $|t-t_{min}|$ $(GeV^2)$ for different photon beam energies. Superscripts are statistical errors and subscripts are systematic errors.}
	\begin{tabular}{|c|c|c|c|c|c|c|}
	\hline\hline
	$E_{\gamma}$ $(GeV)$ & $\sqrt{s}$ $(GeV)$ & \multicolumn{5}{|c|}{$|t-t_{min}|$ $(GeV^2)$} \\
	\cline{3-7}
	 & & $1.660$ & $1.700$ & $1.740$ & $1.780$ & $1.820$ \\
        \hline
      $2.1500$ & $2.2170$ & $0.0382^{\pm0.0163}_{\pm0.0124}$ & $0.0259^{\pm0.0140}_{\pm0.0158}$ & $0.0257^{\pm0.0053}_{\pm0.0159}$ & $0.0196^{\pm0.0047}_{\pm0.0121}$ & - \\
      $2.2500$ & $2.2589$ & $0.0923^{\pm0.0323}_{\pm0.0202}$ & $0.0850^{\pm0.0356}_{\pm0.0216}$ & $0.0835^{\pm0.0449}_{\pm0.0213}$ & $0.0705^{\pm0.0400}_{\pm0.0130}$ & - \\
      $2.3500$ & $2.3001$ & $0.0433^{\pm0.0108}_{\pm0.0096}$ & $0.0562^{\pm0.0145}_{\pm0.0143}$ & $0.0677^{\pm0.0192}_{\pm0.0172}$ & $0.0704^{\pm0.0198}_{\pm0.0130}$ & - \\
      $2.4500$ & $2.3405$ & $0.0401^{\pm0.0074}_{\pm0.0165}$ & $0.0361^{\pm0.0069}_{\pm0.0205}$ & - & - & - \\
      $2.5500$ & $2.3803$ & $0.0347^{\pm0.0076}_{\pm0.0143}$ & $0.0303^{\pm0.0069}_{\pm0.0178}$ & - & - & - \\
      $2.6500$ & $2.4193$ & $0.0306^{\pm0.0180}_{\pm0.0098}$ & $0.0260^{\pm0.0179}_{\pm0.0083}$ & - & - & - \\
      $2.7500$ & $2.4578$ & $0.0212^{\pm0.0133}_{\pm0.0068}$ & $0.0219^{\pm0.0174}_{\pm0.0070}$ & - & - & - \\
	\hline
\hline
\end{tabular}
\label{table:table9}
\end{table*}
\end{center}

\newpage
\clearpage
\section{}
\setcounter{table}{0}
\begin{center}
\begin{table*}[h!]
\captionsetup{width=17cm} 
	\caption{$d\sigma/dt$ $(\mu b/GeV^2)$ vs. $E_{\gamma}$ $(GeV)$ for different photon beam energies. Superscripts are statistical errors and subscripts are systematic errors.}
	\begin{tabular}{|c|c|c|c|c|c|}
	\hline \hline
	$\cos\theta_{cm}$ & \multicolumn{5}{|c|}{$E_{\gamma}$ $(GeV)$} \\
	\cline{2-6}
	 & $1.650$ & $1.750$ & $1.850$ & $1.950$ & $2.050$ \\
	\hline
     $-0.4000$ & - &  $0.2215^{\pm0.0111}_{\pm0.0240}$ & $0.1334^{\pm0.0095}_{\pm0.0242}$ & $0.1212^{\pm0.0091}_{\pm0.0241}$ & $0.0642^{\pm0.0053}_{\pm0.0145}$ \\
     $-0.3000$ & - & $0.1687^{\pm0.0073}_{\pm0.0174}$ & $0.1368^{\pm0.0074}_{\pm0.0217}$ & $0.1136^{\pm0.0075}_{\pm0.0230}$ & $0.0546^{\pm0.0056}_{\pm0.0123}$ \\
     $-0.2000$ & $0.1506^{\pm0.0084}_{\pm0.0242}$ & $0.2013^{\pm0.0080}_{\pm0.0178}$ & $0.1280^{\pm0.0062}_{\pm0.0175}$ & $0.1126^{\pm0.0074}_{\pm0.0174}$ & $0.0528^{\pm0.0034}_{\pm0.0114}$ \\
     $-0.1000$ & $0.1902^{\pm0.0105}_{\pm0.0251}$ & $0.1941^{\pm0.0073}_{\pm0.0178}$ & $0.1361^{\pm0.0062}_{\pm0.0119}$ & $0.1021^{\pm0.0056}_{\pm0.0191}$ & $0.0780^{\pm0.0047}_{\pm0.0155}$ \\
     $0.0000$  & $0.2433^{\pm0.0122}_{\pm0.0275}$ & $0.1979^{\pm0.0073}_{\pm0.0183}$ & $0.1330^{\pm0.0058}_{\pm0.0131}$ & $0.0990^{\pm0.0053}_{\pm0.0081}$ & $0.0894^{\pm0.0054}_{\pm0.0121}$ \\
     $0.1000$  & $0.2102^{\pm0.0103}_{\pm0.0325}$ & $0.2384^{\pm0.0077}_{\pm0.0133}$ & $0.1504^{\pm0.0058}_{\pm0.0095}$ & $0.1112^{\pm0.0051}_{\pm0.0072}$ & $0.0830^{\pm0.0046}_{\pm0.0106}$ \\ 
     $0.2000$  & $0.2396^{\pm0.0114}_{\pm0.0208}$ & $0.2511^{\pm0.0079}_{\pm0.0146}$ & $0.1522^{\pm0.0057}_{\pm0.0083}$ & $0.1312^{\pm0.0057}_{\pm0.0083}$ & $0.1105^{\pm0.0053}_{\pm0.0143}$ \\
     $0.3000$  & $0.2862^{\pm0.0126}_{\pm0.0338}$ & $0.2746^{\pm0.0081}_{\pm0.0173}$ & $0.1805^{\pm0.0062}_{\pm0.0148}$ & $0.1453^{\pm0.0059}_{\pm0.0082}$ & $0.1685^{\pm0.0071}_{\pm0.0177}$ \\ 
     $0.4000$  & $0.2923^{\pm0.0125}_{\pm0.0251}$ & $0.2851^{\pm0.0083}_{\pm0.0132}$ & $0.2402^{\pm0.0075}_{\pm0.0168}$ & $0.2185^{\pm0.0080}_{\pm0.0100}$ & $0.1962^{\pm0.0076}_{\pm0.0112}$ \\
     $0.5000$  & $0.3202^{\pm0.0134}_{\pm0.0176}$ & $0.3126^{\pm0.0087}_{\pm0.0140}$ & $0.2850^{\pm0.0086}_{\pm0.0123}$ & $0.2548^{\pm0.0085}_{\pm0.0155}$ & $0.2576^{\pm0.0090}_{\pm0.0108}$ \\
     $0.6000$  & $0.3226^{\pm0.0129}_{\pm0.0258}$ & $0.3546^{\pm0.0097}_{\pm0.0184}$ & $0.3613^{\pm0.0104}_{\pm0.0257}$ & $0.3270^{\pm0.0100}_{\pm0.0263}$ & $0.3118^{\pm0.0101}_{\pm0.0135}$ \\
     $0.7000$  & $0.3065^{\pm0.0122}_{\pm0.0245}$ & $0.3789^{\pm0.0102}_{\pm0.0268}$ & $0.4184^{\pm0.0111}_{\pm0.0240}$ & $0.4554^{\pm0.0130}_{\pm0.0144}$ & $0.4072^{\pm0.0125}_{\pm0.0224}$ \\
     $0.8000$  & $0.4063^{\pm0.0169}_{\pm0.0298}$ & $0.4961^{\pm0.0134}_{\pm0.0244}$ & $0.5745^{\pm0.0143}_{\pm0.0371}$ & $0.5593^{\pm0.0149}_{\pm0.0258}$ & $0.5167^{\pm0.0277}_{\pm0.0192}$ \\
     $0.9000$  & - & $0.5173^{\pm0.0152}_{\pm0.0311}$ & $0.8667^{\pm0.0231}_{\pm0.0337}$ & $0.7979^{\pm0.0229}_{\pm0.0406}$ & $0.7180^{\pm0.0248}_{\pm0.0474}$ \\
\hline \hline
\end{tabular}
\label{table:wtable1}
\end{table*}
\end{center}

\begin{center}
\begin{table*}[h!]
\captionsetup{width=17cm} 
	\caption{$d\sigma/dt$ $(\mu b/GeV^2)$ vs. $E_{\gamma}$ $(GeV)$ for different photon beam energies. Superscripts are statistical errors and subscripts are systematic errors.}
	\begin{tabular}{|c|c|c|c|c|c|}
	\hline \hline
	$\cos\theta_{cm}$ & \multicolumn{5}{|c|}{$E_{\gamma}$ $(GeV)$} \\
	\cline{2-6}
	 & $2.150$ & $2.250$ & $2.350$ & $2.450$ & $2.550$ \\
	\hline
     $-0.4000$ & $0.0594^{\pm0.0061}_{\pm0.0116}$ & $0.0538^{\pm0.0072}_{\pm0.0101}$ & $0.0386^{\pm0.0069}_{\pm0.0058}$ & $0.0143^{\pm0.0021}_{\pm0.0059}$ & $0.0143^{\pm0.0024}_{\pm0.0062}$ \\
     $-0.3000$ & $0.0408^{\pm0.0039}_{\pm0.0076}$ & $0.0423^{\pm0.0045}_{\pm0.0077}$ & $0.0305^{\pm0.0042}_{\pm0.0053}$ & $0.0125^{\pm0.0017}_{\pm0.0047}$ & $0.0112^{\pm0.0016}_{\pm0.0049}$ \\
     $-0.2000$ & $0.0418^{\pm0.0035}_{\pm0.0081}$ & $0.0418^{\pm0.0042}_{\pm0.0076}$ & $0.0320^{\pm0.0041}_{\pm0.0054}$ & $0.0210^{\pm0.0027}_{\pm0.0051}$ & $0.0201^{\pm0.0026}_{\pm0.0057}$ \\
     $-0.1000$ & $0.0397^{\pm0.0033}_{\pm0.0076}$ & $0.0364^{\pm0.0035}_{\pm0.0077}$ & $0.0335^{\pm0.0039}_{\pm0.0067}$ & $0.0340^{\pm0.0033}_{\pm0.0060}$ & $0.0310^{\pm0.0036}_{\pm0.0054}$ \\
     $0.0000$  & $0.0719^{\pm0.0053}_{\pm0.0111}$ & $0.0539^{\pm0.0048}_{\pm0.0108}$ & $0.0304^{\pm0.0033}_{\pm0.0054}$ & $0.0329^{\pm0.0030}_{\pm0.0078}$ & $0.0242^{\pm0.0028}_{\pm0.0057}$ \\
     $0.1000$ & $0.0841^{\pm0.0056}_{\pm0.0103}$ & $0.0784^{\pm0.0078}_{\pm0.0121}$ & $0.0477^{\pm0.0045}_{\pm0.0081}$ & $0.0379^{\pm0.0032}_{\pm0.0069}$ & $0.0397^{\pm0.0039}_{\pm0.0067}$ \\
     $0.2000$ & $0.1045^{\pm0.0062}_{\pm0.0123}$ & $0.0895^{\pm0.0062}_{\pm0.0113}$ & $0.0805^{\pm0.0071}_{\pm0.0144}$ & $0.0501^{\pm0.0041}_{\pm0.0081}$ & $0.0532^{\pm0.0048}_{\pm0.0077}$ \\
     $0.3000$ & $0.1259^{\pm0.0067}_{\pm0.0078}$ & $0.1141^{\pm0.0062}_{\pm0.0111}$ & $0.0982^{\pm0.0068}_{\pm0.0077}$ & $0.0900^{\pm0.0052}_{\pm0.0113}$ & $0.0745^{\pm0.0063}_{\pm0.0102}$ \\
     $0.4000$ & $0.1539^{\pm0.0073}_{\pm0.0096}$ & $0.1409^{\pm0.0072}_{\pm0.0086}$ & $0.1177^{\pm0.0067}_{\pm0.0096}$ & $0.1220^{\pm0.0063}_{\pm0.0101}$ & $0.1116^{\pm0.0088}_{\pm0.0120}$ \\
     $0.5000$ & $0.2007^{\pm0.0084}_{\pm0.0128}$ & $0.2113^{\pm0.0088}_{\pm0.0121}$ & $0.1526^{\pm0.0070}_{\pm0.0074}$ & $0.1694^{\pm0.0075}_{\pm0.0105}$ & $0.1486^{\pm0.0106}_{\pm0.0100}$ \\
     $0.6000$ & $0.2713^{\pm0.0093}_{\pm0.0102}$ & $0.2044^{\pm0.0080}_{\pm0.0099}$ & $0.2369^{\pm0.0104}_{\pm0.0153}$ & $0.2206^{\pm0.0099}_{\pm0.0154}$ & $0.2083^{\pm0.0150}_{\pm0.0150}$ \\
     $0.7000$ & $0.3208^{\pm0.0110}_{\pm0.0141}$ & $0.3082^{\pm0.0115}_{\pm0.0166}$ & $0.3011^{\pm0.0112}_{\pm0.0134}$ & $0.2787^{\pm0.0117}_{\pm0.0128}$ & $0.2656^{\pm0.0189}_{\pm0.0205}$ \\
     $0.8000$ & $0.4965^{\pm0.0165}_{\pm0.0215}$ & $0.4433^{\pm0.0153}_{\pm0.0260}$ & $0.4442^{\pm0.0183}_{\pm0.0233}$ & $0.3782^{\pm0.0150}_{\pm0.0160}$ & $0.3676^{\pm0.0267}_{\pm0.0257}$ \\
     $0.9000$ & $0.7900^{\pm0.0323}_{\pm0.0435}$ & $0.6411^{\pm0.0261}_{\pm0.0266}$ & $0.5566^{\pm0.0279}_{\pm0.0306}$ & $0.4633^{\pm0.0225}_{\pm0.0238}$ & $0.5053^{\pm0.0494}_{\pm0.0244}$ \\
\hline \hline
\end{tabular}
\label{table:wtable2}
\end{table*}
\end{center}

\begin{center}
\begin{table*}[h!]
\captionsetup{width=17cm} 
	\caption{$d\sigma/dt$ $(\mu b/GeV^2)$ vs. $E_{\gamma}$ $(GeV)$ for different photon beam energies. Superscripts are statistical errors and subscripts are systematic errors.}
	\begin{tabular}{|c|c|c|c|c|c|}
	\hline \hline
	$\cos\theta_{cm}$ & \multicolumn{5}{|c|}{$E_{\gamma}$ $(GeV)$} \\
	\cline{2-6}
	 & $2.650$ & $2.750$ & $2.850$ & $2.950$ & $3.050$ \\
	\hline
     $-0.4000$ & $0.0079^{\pm0.0057}_{\pm0.0056}$ & $0.0067^{\pm0.0013}_{\pm0.0049}$ & - & - & - \\
     $-0.3000$ & $0.0135^{\pm0.0021}_{\pm0.0097}$ & $0.0136^{\pm0.0064}_{\pm0.0093}$ & - & - & - \\
     $-0.2000$ & $0.0153^{\pm0.0021}_{\pm0.0107}$ & $0.0128^{\pm0.0022}_{\pm0.0089}$ & - & - & - \\
     $-0.1000$ & $0.0129^{\pm0.0018}_{\pm0.0052}$ & $0.0095^{\pm0.0018}_{\pm0.0040}$ & $0.0219^{\pm0.0027}_{\pm0.0139}$ & $0.0137^{\pm0.0023}_{\pm0.0076}$ & $0.0066^{\pm0.0034}_{\pm0.0057}$ \\
     $0.0000$  & $0.0184^{\pm0.0022}_{\pm0.0081}$ & $0.0116^{\pm0.0017}_{\pm0.0048}$ & $0.0225^{\pm0.0032}_{\pm0.0082}$ & $0.0157^{\pm0.0027}_{\pm0.0058}$ & $0.0074^{\pm0.0017}_{\pm0.0055}$ \\
     $0.1000$  & $0.0233^{\pm0.0027}_{\pm0.0056}$ & $0.0241^{\pm0.0028}_{\pm0.0061}$ & $0.0312^{\pm0.0036}_{\pm0.0082}$ & $0.0243^{\pm0.0035}_{\pm0.0069}$ & $0.0117^{\pm0.0022}_{\pm0.0075}$ \\
     $0.2000$  & $0.0394^{\pm0.0039}_{\pm0.0053}$ & $0.0280^{\pm0.0035}_{\pm0.0037}$ & $0.0328^{\pm0.0038}_{\pm0.0069}$ & $0.0254^{\pm0.0034}_{\pm0.0064}$ & $0.0138^{\pm0.0024}_{\pm0.0058}$ \\
     $0.3000$ &  $0.0494^{\pm0.0044}_{\pm0.0060}$ & $0.0382^{\pm0.0043}_{\pm0.0046}$ & $0.0351^{\pm0.0041}_{\pm0.0051}$ & $0.0287^{\pm0.0039}_{\pm0.0040}$ & $0.0474^{\pm0.0061}_{\pm0.0086}$ \\
     $0.4000$  & $0.0633^{\pm0.0054}_{\pm0.0058}$ & $0.0659^{\pm0.0062}_{\pm0.0061}$ & $0.0468^{\pm0.0050}_{\pm0.0058}$ & $0.0422^{\pm0.0052}_{\pm0.0053}$ & $0.0444^{\pm0.0057}_{\pm0.0093}$ \\
     $0.5000$  & $0.0928^{\pm0.0070}_{\pm0.0079}$ & $0.0833^{\pm0.0076}_{\pm0.0071}$ & $0.0687^{\pm0.0070}_{\pm0.0068}$ & $0.0660^{\pm0.0077}_{\pm0.0063}$ & $0.0422^{\pm0.0058}_{\pm0.0082}$ \\
     $0.6000$  & $0.1327^{\pm0.0100}_{\pm0.0083}$ & $0.1222^{\pm0.0102}_{\pm0.0077}$ & $0.1313^{\pm0.0117}_{\pm0.0096}$ & $0.1159^{\pm0.0116}_{\pm0.0082}$ & $0.1083^{\pm0.0120}_{\pm0.0127}$ \\
     $0.7000$  & $0.2510^{\pm0.0195}_{\pm0.0129}$ & $0.2239^{\pm0.0181}_{\pm0.0123}$ & $0.1846^{\pm0.0160}_{\pm0.0095}$ & $0.2089^{\pm0.0192}_{\pm0.0143}$ & $0.1637^{\pm0.0178}_{\pm0.0130}$ \\
     $0.8000$  & $0.3070^{\pm0.0254}_{\pm0.0140}$ & $0.3217^{\pm0.0276}_{\pm0.0153}$ & $0.3215^{\pm0.0312}_{\pm0.0158}$ & $0.2974^{\pm0.0331}_{\pm0.0148}$ & $0.3350^{\pm0.0395}_{\pm0.0207}$ \\
     $0.9000$  & $0.5552^{\pm0.0634}_{\pm0.0290}$ & $0.5942^{\pm0.0795}_{\pm0.0322}$ & $0.6169^{\pm0.0850}_{\pm0.0358}$ & $0.6322^{\pm0.1151}_{\pm0.0426}$ & $0.6110^{\pm0.1237}_{\pm0.0538}$ \\
\hline \hline
\end{tabular}
\label{table:wtable3}
\end{table*}
\end{center}

\begin{center}
\begin{table*}[h!]
\captionsetup{width=17cm} 
	\caption{$d\sigma/dt$ $(\mu b/GeV^2)$ vs. $E_{\gamma}$ $(GeV)$ for different photon beam energies. Superscripts are statistical errors and subscripts are systematic errors.}
	\begin{tabular}{|c|c|c|c|c|c|}
	\hline \hline
	$\cos\theta_{cm}$ & \multicolumn{5}{|c|}{$E_{\gamma}$ $(GeV)$} \\
	\cline{2-6}
	 & $3.150$ & $3.250$ & $3.350$ & $3.450$ & $3.550$ \\
	\hline
     $-0.1000$ & $0.0054^{\pm0.0017}_{\pm0.0043}$ & $0.0039^{\pm0.0012}_{\pm0.0032}$ & $0.0041^{\pm0.0017}_{\pm0.0034}$ & $0.0042^{\pm0.0018}_{\pm0.0038}$ & $0.0022^{\pm0.0012}_{\pm0.0020}$ \\
     $0.0000$  & $0.0069^{\pm0.0023}_{\pm0.0048}$ & $0.0026^{\pm0.0017}_{\pm0.0023}$ & $0.0028^{\pm0.0010}_{\pm0.0022}$ & $0.0024^{\pm0.0011}_{\pm0.0023}$ & $0.0024^{\pm0.0015}_{\pm0.0022}$ \\
     $0.1000$  & $0.0094^{\pm0.0024}_{\pm0.0057}$ & $0.0061^{\pm0.0015}_{\pm0.0038}$ & $0.0075^{\pm0.0020}_{\pm0.0043}$ & $0.0050^{\pm0.0024}_{\pm0.0043}$ & $0.0041^{\pm0.0035}_{\pm0.0036}$ \\
     $0.2000$  & $0.0174^{\pm0.0038}_{\pm0.0084}$ & $0.0086^{\pm0.0019}_{\pm0.0044}$ & $0.0072^{\pm0.0019}_{\pm0.0035}$ & $0.0073^{\pm0.0022}_{\pm0.0055}$ & $0.0055^{\pm0.0026}_{\pm0.0044}$ \\
     $0.3000$  & $0.0387^{\pm0.0060}_{\pm0.0080}$ & $0.0179^{\pm0.0033}_{\pm0.0061}$ & $0.0130^{\pm0.0027}_{\pm0.0055}$ & $0.0252^{\pm0.0046}_{\pm0.0105}$ & $0.0131^{\pm0.0037}_{\pm0.0081}$ \\
     $0.4000$  & $0.0414^{\pm0.0069}_{\pm0.0087}$ & $0.0301^{\pm0.0048}_{\pm0.0073}$ & $0.0294^{\pm0.0049}_{\pm0.0107}$ & $0.0140^{\pm0.0031}_{\pm0.0084}$ & $0.0124^{\pm0.0034}_{\pm0.0064}$ \\
     $0.5000$  & $0.0463^{\pm0.0074}_{\pm0.0090}$ & $0.0448^{\pm0.0067}_{\pm0.0103}$ & $0.0441^{\pm0.0076}_{\pm0.0111}$ & $0.0394^{\pm0.0069}_{\pm0.0108}$ & $0.0318^{\pm0.0072}_{\pm0.0118}$ \\
     $0.6000$  & $0.1145^{\pm0.0152}_{\pm0.0152}$ & $0.0868^{\pm0.0117}_{\pm0.0111}$ & $0.0768^{\pm0.0111}_{\pm0.0098}$ & $0.0823^{\pm0.0133}_{\pm0.0115}$ & $0.0538^{\pm0.0111}_{\pm0.0099}$ \\
     $0.7000$  & $0.1708^{\pm0.0225}_{\pm0.0125}$ & $0.2053^{\pm0.0266}_{\pm0.0160}$ & $0.1952^{\pm0.0276}_{\pm0.0172}$ & $0.1762^{\pm0.0270}_{\pm0.0188}$ & $0.1219^{\pm0.0219}_{\pm0.0132}$ \\
     $0.8000$  & $0.3292^{\pm0.0444}_{\pm0.0200}$ & $0.3166^{\pm0.0462}_{\pm0.0228}$ & $0.2496^{\pm0.0429}_{\pm0.0187}$ & $0.2298^{\pm0.0481}_{\pm0.0233}$ & $0.1788^{\pm0.0395}_{\pm0.0184}$ \\
     $0.9000$  & $0.6351^{\pm0.1402}_{\pm0.0577}$ & $0.5407^{\pm0.1414}_{\pm0.0545}$ & $0.5969^{\pm0.2243}_{\pm0.0611}$ & $0.5216^{\pm0.1725}_{\pm0.0945}$ & $0.3243^{\pm0.1841}_{\pm0.0572}$ \\
\hline \hline
\end{tabular}
\label{table:wtable4}
\end{table*}
\end{center}

\newpage
\clearpage
\section{}
\setcounter{table}{0}
\begin{center}
\begin{table*}[h!]
\captionsetup{width=17cm} 
	\caption{$d\sigma/dt(|t|=|t|_{min})$ $(\mu b/GeV^2)$ and $B$ $(GeV^{-2})$ vs. $E_{\gamma}$ $(GeV)$ photon beam energy. Superscripts are statistical errors and subscripts are systematic errors.}
	\begin{tabular}{|c|c|c|c|}
	\hline \hline
	$E_{\gamma}$ $(GeV)$ & $\sqrt{s}$ $(GeV)$ & $d\sigma/dt(|t|=|t|_{min})$ $(\mu b/GeV^2)$ & \hspace{0.5cm} $B$ $(GeV^{-2})$ \hspace{0.5cm} \\
	\hline
    $1.6500$ & $1.9942$ & $0.4015^{\pm0.0170}_{\pm0.0156}$ &   $3.1636^{\pm0.3334}_{\pm0.1249}$ \\
    $1.7500$ & $2.0407$ & $0.5605^{\pm0.0280}_{\pm0.0327}$ &   $3.1036^{\pm0.3358}_{\pm0.1742}$ \\
    $1.8500$ & $2.0861$ & $0.6796^{\pm0.0206}_{\pm0.0599}$ &   $2.8647^{\pm0.1548}_{\pm0.3436}$ \\
    $1.9500$ & $2.1306$ & $0.9525^{\pm0.0219}_{\pm0.0725}$ &   $3.8019^{\pm0.0964}_{\pm0.2544}$ \\
    $2.0500$ & $2.1742$ & $0.9088^{\pm0.0353}_{\pm0.1082}$ &   $3.5221^{\pm0.1557}_{\pm0.4297}$ \\
    $2.1500$ & $2.2170$ & $0.8779^{\pm0.0229}_{\pm0.0609}$ &   $3.1348^{\pm0.0866}_{\pm0.2139}$ \\
    $2.2500$ & $2.2589$ & $0.7138^{\pm0.0300}_{\pm0.0448}$ &   $2.8456^{\pm0.1232}_{\pm0.1924}$ \\
    $2.3500$ & $2.3001$ & $0.5735^{\pm0.0304}_{\pm0.0357}$ &   $2.2987^{\pm0.1429}_{\pm0.2173}$ \\
    $2.4500$ & $2.3405$ & $0.7866^{\pm0.0230}_{\pm0.0383}$ &   $2.2048^{\pm0.0659}_{\pm0.1144}$ \\
    $2.5500$ & $2.3803$ & $0.7069^{\pm0.0559}_{\pm0.0545}$ &   $2.6593^{\pm0.1549}_{\pm0.1585}$ \\
    $2.6500$ & $2.4193$ & $0.9782^{\pm0.1859}_{\pm0.0608}$ &   $3.0617^{\pm0.3422}_{\pm0.1524}$ \\
    $2.7500$ & $2.4578$ & $0.0389^{\pm0.1518}_{\pm0.0582}$ &   $3.1467^{\pm0.2874}_{\pm0.1416}$ \\
    $2.8500$ & $2.4957$ & $1.1070^{\pm0.1065}_{\pm0.0540}$ &   $3.4009^{\pm0.2001}_{\pm0.1412}$ \\
    $2.9500$ & $2.5330$ & $1.1376^{\pm0.1309}_{\pm0.0814}$ &   $3.2025^{\pm0.2366}_{\pm0.1498}$ \\
    $3.0500$ & $2.5698$ & $1.5404^{\pm0.2722}_{\pm0.1235}$ &   $3.8577^{\pm0.3594}_{\pm0.1835}$ \\
    $3.1500$ & $2.6061$ & $1.1367^{\pm0.1662}_{\pm0.1220}$ &   $3.1332^{\pm0.2685}_{\pm0.1820}$ \\
    $3.2500$ & $2.6418$ & $1.3291^{\pm0.2542}_{\pm0.1479}$ &   $3.1422^{\pm0.3412}_{\pm0.1903}$ \\
    $3.3500$ & $2.6771$ & $1.5029^{\pm0.5328}_{\pm0.2504}$ &   $3.2923^{\pm0.5067}_{\pm0.4077}$ \\
    $3.4500$ & $2.7119$ & $1.7548^{\pm0.5939}_{\pm0.2266}$ &   $3.3680^{\pm0.4705}_{\pm0.3440}$ \\
    $3.5500$ & $2.7463$ & $0.8867^{\pm0.2394}_{\pm0.2558}$ &   $2.8285^{\pm0.3924}_{\pm0.4753}$ \\
\hline \hline
\end{tabular}
\label{table:table1}
\end{table*}
\end{center}

\clearpage
\twocolumngrid

\end{document}